\documentclass[twocolumn]{aastex63_mod}

\usepackage{xspace}
\usepackage{amsmath}
\usepackage{hyperref}
\usepackage{makecell}

\newcommand{\msyr}{\hbox{$M_\odot~\text{yr}^{-1}$}\xspace}

\newcommand{\um}{\hbox{$\mu$m}\xspace}
\newcommand{\Ha}{H$\alpha$\xspace}
\newcommand{\Hb}{H$\beta$\xspace}
\newcommand{\OIII}{[\ion{O}{3}] $\lambda 5007$\xspace}
\newcommand{\OIIIsimp}{[\ion{O}{3}] \xspace}
\newcommand{\NIIsimp}{[\ion{N}{2}]}

\newcommand{\NIIt}{[\ion{N}{2}] $\lambda \lambda 6548,~6584$\xspace}
\newcommand{\NII}{[\ion{N}{2}] $\lambda 6584$\xspace}
\newcommand{\NIIl}{[\ion{N}{2}] $\lambda 6548$\xspace}
\newcommand{\jhmag}{\hbox{$m_{J+JH+H}$}\xspace}
\newcommand{\umin}{\ensuremath{U_{\rm min}}\xspace}

\newcommand{\gam}{\ensuremath{\gamma}\xspace}

\newcommand{\mdust}{\ensuremath{M_{\rm dust}}\xspace}

\newcommand{\mcsed}{\texttt{MCSED}\xspace}
\newcommand{\cloudy}{C{\footnotesize LOUDY}\xspace}
\usepackage{booktabs}
\usepackage{changepage}

\date{\today}
\shorttitle{Star Forming Galaxies at $1.2<z<1.9$}
\shortauthors{Nagaraj, Ciardullo, Bowman, Gronwall}
\graphicspath{{./}{figures/}}

\begin{document}

\title{Stellar, Gas, and Dust Emission of Star Forming Galaxies out to $z\sim2$}

\correspondingauthor{Gautam Nagaraj}
\email{gxn75@psu.edu}

\author[0000-0002-0905-342X]{Gautam Nagaraj}
\affiliation{Department of Astronomy \& Astrophysics and Institute for Gravitation and the Cosmos, The Pennsylvania
State University, University Park, PA 16802}

\author[0000-0002-1328-0211]{Robin Ciardullo}
\affiliation{Department of Astronomy \& Astrophysics and Institute for Gravitation and the Cosmos, The Pennsylvania
State University, University Park, PA 16802}

\author[0000-0003-4381-5245]{William P. Bowman}
\affiliation{Department of Astronomy \& Astrophysics and Institute for Gravitation and the Cosmos, The Pennsylvania
State University, University Park, PA 16802}

\author[0000-0001-6842-2371]{Caryl Gronwall}
\affiliation{Department of Astronomy \& Astrophysics and Institute for Gravitation and the Cosmos, The Pennsylvania
State University, University Park, PA 16802}

\begin{abstract}

While dust is a major player in galaxy evolution, its relationship with gas and stellar radiation in the early universe is still not well understood. We combine 3D-HST emission line fluxes with far-UV through far-IR photometry in a sample of 669 emission-line galaxies (ELGs) between $1.2 < z < 1.9$ and use the \mcsed spectral energy distribution fitting code to constrain the galaxies' physical parameters, such as their star formation rates (SFRs), stellar masses, and dust masses. We find that the assumption of energy balance between dust attenuation and emission is likely unreasonable in many cases. We highlight a relationship between the mass-specific star formation rate (sSFR), stellar mass, and dust mass, although its exact form is still unclear. Finally, a stacking of \Ha and \Hb fluxes shows that nebular attenuation increases with stellar mass and SFR for IR-bright ELGs.

\end{abstract}


\section{Introduction} \label{sec:intro}

To better understand galaxy formation and evolution, we must unravel the interplay between gas, stars, dark matter, and dust in different environments over a wide redshift range. Although dust constitutes a small fraction of the interstellar medium (ISM\null) by mass, it plays a vital role in star formation, ISM evolution, and chemistry \citep[e.g.,][]{Draine2003rev,Draine2003}, and is responsible for up to 30\% of a galaxy’s bolometric luminosity via the reprocessing of absorbed radiation \citep{Bernstein2002}.

In the local universe, \cite{Draine2003} and \cite{Zubko2004} found that dust likely consists of silicates and carbonaceous grains, with polycyclic aromatic hydrocarbons (PAHs) potentially containing up to 15-20\% of the carbon in the ISM. Photoelectrons from these PAHs are among the most important heating mechanisms for photon-dominated regions \citep{Tielens1985}. In addition, PAH molecules and dust grains serve as catalysts for chemical reactions that create the variety of neutral and charged molecules observed in the ISM \citep[e.g.,][]{Galliano2008}. 

\cite{Engelbracht2005} and \cite{Madden2006}, among others, have found that in the local universe, PAH emission correlates with gas-phase metallicity, which supports the astrophysical picture that metals are formed by stellar nucleosynthesis and supernovae, with a fraction of the metals being frozen out in dust grains. At high redshift, however, most stars would not have yet been able to evolve to the asymptotic giant branch (AGB) stage, which is when most PAHs are probably created \citep{Dwek1998}. Furthermore, the lack of Type Ia supernovae with respect to core collapse and pair-instability supernovae would affect the ratio of iron-peak to $\alpha$-process elements \citep{Greggio1983}, thus changing the fraction of metals that get locked in dust grains. Therefore, in the early universe, we have reasons to suspect that dust, gas, and radiation interact differently on the scale of galaxies.

While the relationships between dust, gas, and stars have been widely studied in the local universe \citep[e.g.,][]{Dunne2000,Draine2007,daCunha2008,Galliano2008}, the study of their evolution through cosmic time has been stymied by the lack of comprehensive photometric and spectroscopic observations (rest-frame optical through mid- and far-IR) at redshifts beyond $z \sim 0.5$.  In fact, the details of the reprocessing of starlight by dust constitute one of the most significant sources of uncertainty in measurements of star formation rates in the early universe \citep[e.g.,][]{Bouwens2012,Finkelstein2012,Oesch2013}.

Many of the studies that have been conducted on the properties of dust emission and absorption in the high-redshift universe have been straightforward analyses using samples of galaxies selected via their high star formation rates (SFRs) or specific SFRs (sSFRs). For example, a number of studies have investigated how the strength of the 2175~\AA~  absorption feature varies with SFR, galaxy color, and other parameters \citep[e.g.,][]{Noll2009,Conroy2010Dust,Buat2012}.  Others have evaluated the performance of different star formation indicators \citep[e.g.,][]{Utomo2014} and analyzed the relationship between the attenuation of UV light from O and B stars and the re-radiated IR luminosity from dust, otherwise known as the IRX-$\beta$ relation \citep[e.g.,][]{Meurer1999, Reddy2012}. While these investigations have begun to illuminate the interplay among dust, gas, and radiation at high redshift, they often suffer from the use of incomplete, biased samples and the lack of spectroscopic information. As detailed in \cite{Nagaraj2021}, hereafter referred to as Paper~I, while emission line galaxy (ELG) surveys are subject to their own forms of incompleteness and bias (e.g., a minimum SFR is required to produce detectable emission lines), they provide more accurate redshifts and comprise a diverse range of stellar masses, metallicities, and SFRs \citep[e.g.,][]{Momcheva2016,GrasshornGebhardt2016,Bowman2019}.

In Paper~I, we vetted a clean sample of $1.16<z<1.90$ ELGs from the 3D-HST Treasury program \citep{Brammer2012,Momcheva2016}, a survey of the five CANDELS fields using the \textit{Hubble Space Telescope} WFC3 G141 grism.  These galaxies have $m_{JH} < 26$ and unambiguous redshifts based on the distinctive shape of their \OIIIsimp $\lambda\lambda 4959,5007$ doublet (all redshifts) and/or the presence of \Ha ($z\lesssim 1.5$) emission. By combining the emission line strengths and trustworthy grism redshifts with the comprehensive photometric data available for the 3D-HST/CANDELS fields, we measured the galaxies' stellar masses, SFRs, and internal extinction via their rest-frame UV through near-IR spectral energy distributions (SEDs).  In this paper, we build upon the study by \cite{Puglisi2016} and add mid- and far-IR photometry to the data set, to address a fundamental question relating dust attenuation to dust emissivity: the issue of energy balance.

The idea of energy balance between stellar radiation absorbed by dust and the thermal re-radiation of this energy in the mid- and far-infrared is highly embedded in astronomical theory. In practice, though, the energy balance relationship may not be entirely straightforward. Dust attenuation is dependent on the star-dust geometry. Two galaxies with the same total dust mass can have very different overall attenuation if, for example, one has a clumpy dust distribution while the other has its dust smoothly distributed. Therefore, even if energy balance is physically accurate, measuring the attenuation of the UV light may not guarantee a proper calibration of dust IR emission. 

Energy balance between attenuated UV/optical light and IR emission is implicitly encoded into the IRX-$\beta$ relation, which connects a star-forming galaxy's UV slope ($\beta$) to its infrared excess \citep[e.g.,][]{ Overzier2011,Hao2011,Buat2012}. Although such a correlation has been shown to exist \citep{Meurer1999,Hao2011}, the scatter around the relation is often over an order of magnitude \citep[e.g.,][]{Reddy2012,Buat2012}. \cite{Narayanan2018} have used cosmological simulations to investigate the physical mechanisms that give rise to this scatter; we can use the physical insights from this model to estimate the configuration of matter in IR-bright ELGs in our redshift range.

In \S \ref{sec:data} of this paper, we describe the data that we use to investigate the properties of dust, gas, and radiation at redshifts $1.16<z<1.90$. In \S \ref{sec:methods}, we briefly summarize our selection of a clean sample of emission-line galaxies with unambiguous redshifts, the removal of active galactic nuclei from the data (AGN), the incorporation of mid- and far-IR photometry into the galaxy SEDs, and our SED-fitting analysis.  In \S \ref{sec:results}, we present and discuss results that demonstrate the effect that the energy balance assumption has on the anti-correlation between UV slope and long wavelength emission.  Finally, in \S \ref{sec:disc}, we summarize the investigation and discuss implications, limitations, and future directions. We assume a $\Lambda$CDM cosmology with $\Omega_{\Lambda} = 0.69$, $\Omega_M = 0.31$ and $H_0 = 69$~km~s$^{-1}$~Mpc$^{-1}$ \citep{Bennett2013}. All magnitudes in the paper use the AB magnitude system \citep{Oke1974}.

This paper is the second of three papers dealing with the properties of ELGs at redshifts $1.2<z<1.9$. Paper~I describes the construction and vetting of our sample and presents correlations among stellar mass and various observational and derived properties. Paper~III (Nagaraj et al.~in prep) will give the \Ha and \OIII luminosity functions for our clean sample and measure the bias of the emission-line galaxies, which will be a very useful for determining how well we will be able to measure cosmological parameters in upcoming IR-grism missions such as \textit{Euclid} and the
\textit{Nancy Grace Roman Space Telescope} (\textit{NGRST}) missions.

\section{Data}\label{sec:data}

\citet{Bowman2019} assembled a vetted sample of 1952 ELGs from the 3D-HST survey \citep[GO-11600, 12177, 12328;][]{Brammer2012, Momcheva2016} with unambiguous emission-line redshifts between $1.90 < z < 2.35$ and near-IR continuum magnitudes $m_{JH} \leq 26$.  In Paper~I, we extended this work by creating a similarly vetted sample of 4,350 galaxies in the redshift range $1.2 < z < 1.9$.  Taken together, the two surveys catalog all $m_{JH} \leq 26$ sources with \OIII present in the WFC3 G141 grism frames.  

Because the 3D-HST survey was conducted within the five CANDELS fields \citep{Grogin2011,Koekemoer2011}, all of our galaxies have extensive rest-frame UV through IR imaging from both ground- and space-based telescopes.  \cite{Skelton2014} carefully combined these measurements into a comprehensive set of point spread function (PSF)-matched photometry extending from 0.3 to 8 \um (observed frame). In Paper~I, we appended near-UV photometry to this dataset using photometry from the Deep Imaging Survey of \textit{GALEX} \citep{Martin2005, Morrissey2007} and the deep images taken of GOODS-S using the \textit{Swift}/UVOT instrument \citep{Hoversten2009}.

Here we add mid- and far-IR photometry to that dataset.  In the redshift range $1.2 < z < 1.9$, about 15\% (669/4,350) of the ELGs have been detected with the \textit{Spitzer} Multiband Imaging Photometer \citep[MIPS, 24+70+160 \um;][]{Rieke2004}, the \textit{Herschel} Photodetector Array Camera and Spectrometer \citep[PACS, 70+100+160 \um;][]{Poglitsch2010}, and/or the \textit{Herschel} Spectral and Photometric Imaging Receiver \citep[SPIRE, 250+350+500 \um;][]{Griffin2010}. The CANDELS team has carefully collected this mid- and far-IR photometry and identified their most likely optical/NIR counterparts on \textit{Hubble} F160W and \textit{Spitzer} IRAC images \citep{Barro2019}. In this study, we use the catalogued 24 and 70~\um photometry from MIPS (full width half maxima (FWHMs) of 6\arcsec and 18\arcsec), the 100 and 160~\um measurements of PACS (FWHM of 7\arcsec and 11\arcsec), and the 250 and 350~\um data of SPIRE (FWHM of 18\farcs 1 and 25\farcs 2).  Table \ref{tab:IRNumbers} shows the number of measured fluxes in each photometric band. We will present the full catalog in Paper~III.

\begin{figure*}
    \centering
    \resizebox{\hsize}{!}{
    \includegraphics{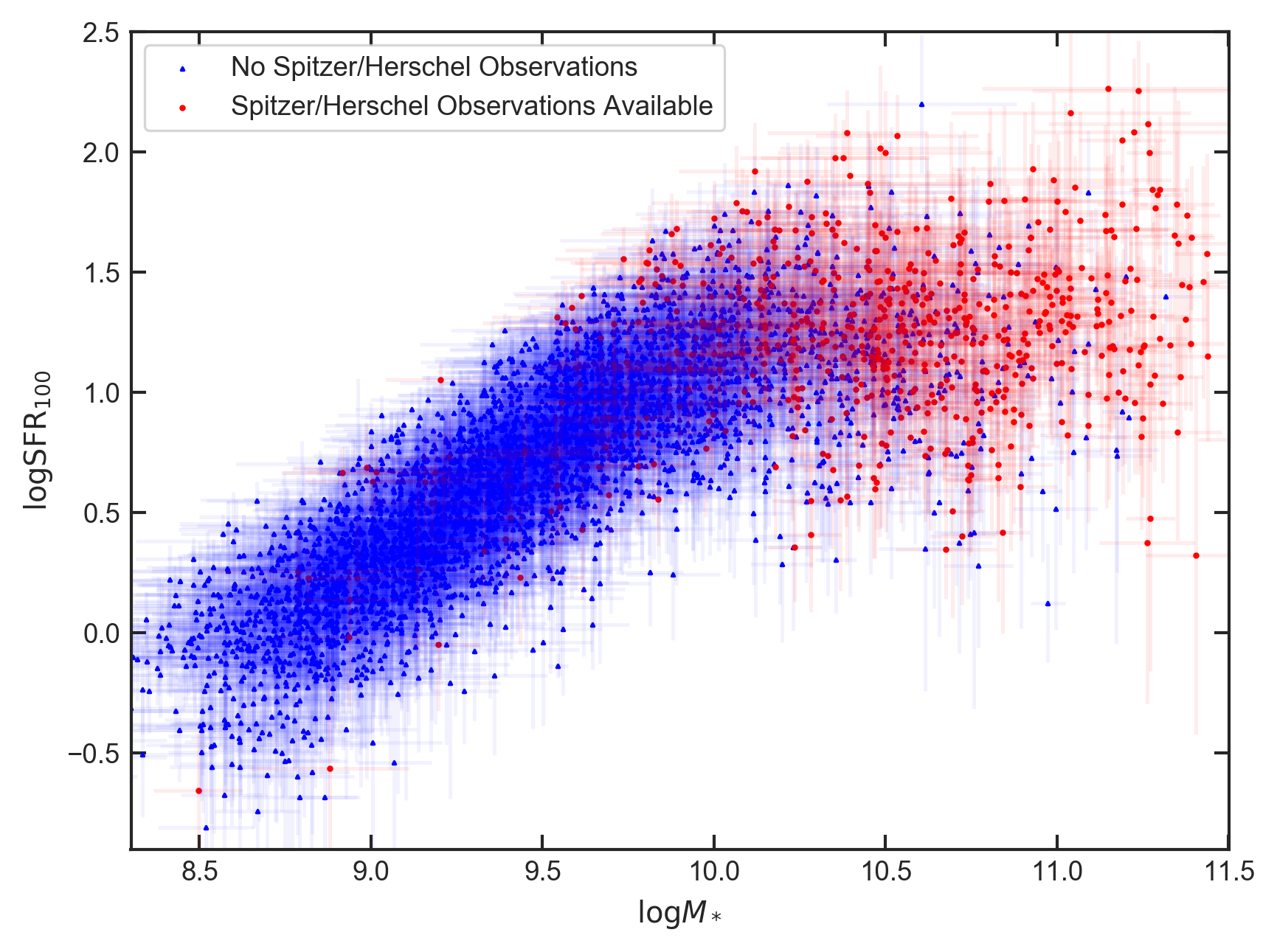}
    \includegraphics{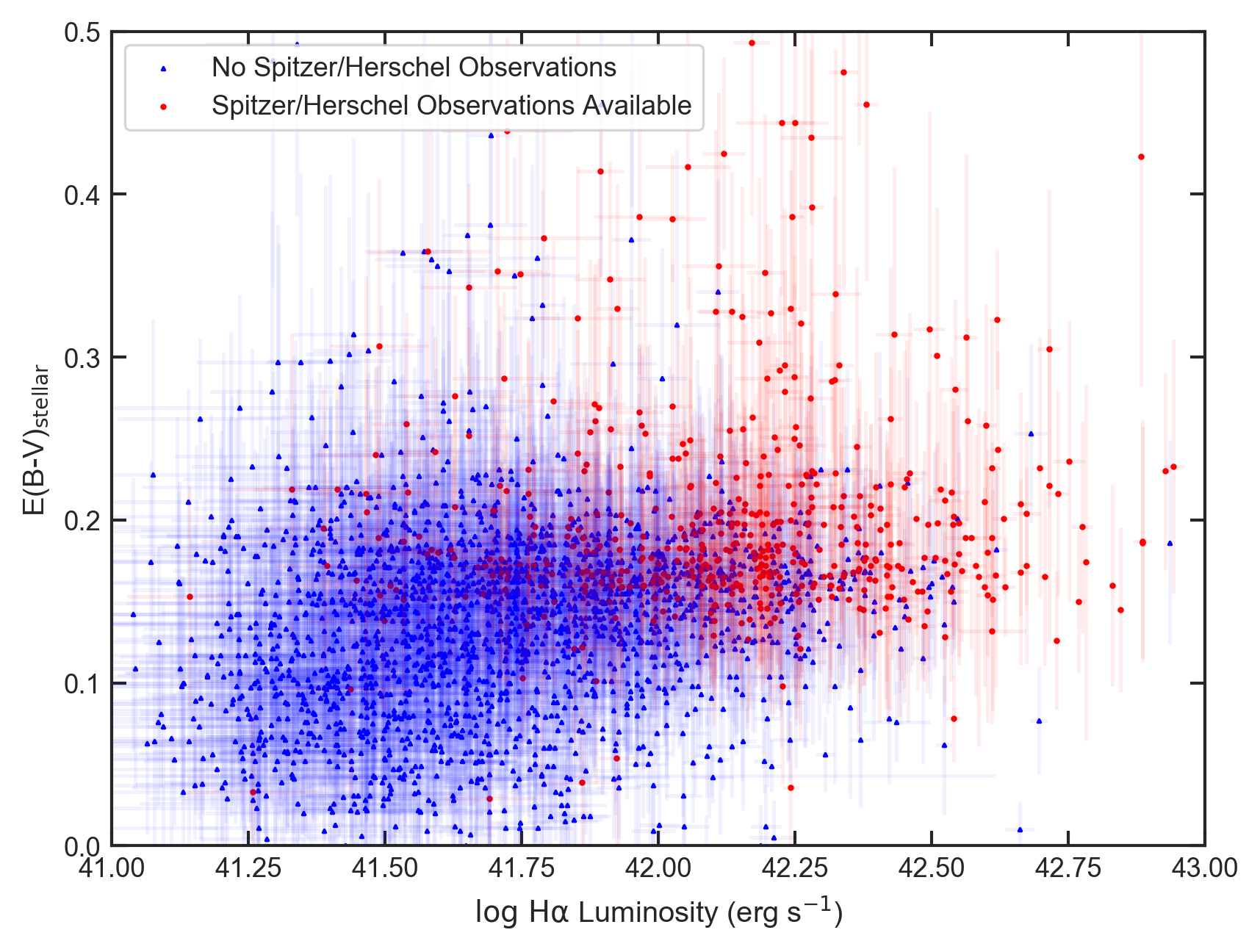}}
    \caption{Properties of our IR sample of ELGs compared to the ELGs that are not detected in the mid- or far-IR\null.  From the left panel, we observe that the IR sample has a much higher average mass and a somewhat higher average SFR (but with many quiescent and moderately-star-forming galaxies). From the right panel, we find that the IR sample features more dust attenuation and larger \Ha luminosities.}
    \label{fig:sampleprop}
\end{figure*}

This 669-object sample with \textit{Spitzer} and/or \textit{Herschel} measurements, which we hereafter call the ``IR-bright'' or simply ``IR'' sample, has noticeably different properties from the 4,350-member parent sample.  This is expected, given the requirement that these objects be detected in MIR/FIR survey that have relatively bright flux limits.  In Figure \ref{fig:sampleprop}, we show the distributions of stellar mass vs.\ SFR and stellar E(B-V) vs.\ \Ha luminosity (for sources $1.16<z<1.56$) for the IR sample and for the non-IR sample. To make a fair comparison, the stellar masses, SFRs, and E(B-V) values are inferred with identical assumptions, and only UV-NIR spectra are used in the fit.  (See \S \ref{subsec:individual} for more details.)

From the figure, it is clear that the galaxies of the IR sample tend to have higher stellar masses, attenuation, and H$\alpha$ luminosities as well as somewhat higher SFRs (though a number of quiescent or moderately-star-forming objects are included in the sample). These effects generally are associated with the mass bias. We explore the impacts of these biases in \S \ref{sec:results}.

Given the much larger PSF of the mid- and far-IR instruments, source confusion is potentially a major problem. \cite{Barro2019} employed various measures to reduce the effects of confusion noise in their catalog, with the most important one being a careful matching of sources from better to worse angular resolution. Moreover, we find that the numbers of F160W sources within the full width half maximum (FWHM) of the various IR images, as catalogued by \cite{Barro2019}, do not correlate with any physical or observational property, such as apparent JH magnitude or dust mass. For this reason, we have adopted the \cite{Barro2019} flux values without modification for our analyses. However, we do caution that source confusion may still be present and cause slight systematic shifts in our results, albeit in a random or unpredictable fashion. 

\begin{table}[]
\hspace*{-1.8cm}\begin{tabular}{@{}lcccccc@{}}
\toprule
Telescope           & \multicolumn{2}{c}{\textit{Spitzer}} & \multicolumn{4}{c}{\textit{Herschel}}                \\ \hline
Instrument          & \multicolumn{2}{c}{MIPS}             & \multicolumn{2}{c}{PACS} & \multicolumn{2}{c}{SPIRE} \\ \midrule
Wavelength ($\mu$m) & 24                & 70               & 100         & 160        & 250         & 350         \\ 
Source Count        & 568               & 29               & 262         & 271        & 103         & 84          \\ \bottomrule
\end{tabular}
\caption{Number of $1.2 < z < 1.9$ sources in our vetted 3D-HST catalog with rest-frame mid- and far-IR data.  Of this sample, most (568/669) have MIPS 24 \um data whereas almost none have MIPS 70 \um data. Under half of the sources have PACS 100 and/or 160 \um data. Around 13\% of sources have SPIRE 250 and/or 350 \um data.}
\label{tab:IRNumbers}
\end{table}

\section{Methods}\label{sec:methods}

Using a set of five quality indices (overall trust in redshift determination, emission line prominence, fullness of grism image, presence of continuum-like contamination, and presence of line-like contamination), we assembled a sample of 4,350 trustworthy ELGs from a pool of 9,341 candidates in the 3D-HST database. As our analyses are geared toward normal star-forming galaxies, we removed active galactic nuclei (AGN) from the sample by cross-correlating 3D-HST sources with \textit{Chandra} X-ray catalogs of the CANDELS fields \citep{Nandra2015,Civano2016,Luo2017, Kocevski2018,Suh2019}. Out of the 4,350 objects originally identified, 72 (1.7\%) were within 1\arcsec~of an X-ray source with $\log L_X>42$~ergs~s$^{-1}$ and therefore classified as AGN\null. 

Stacked X-ray analysis revealed background X-ray levels consistent with those expected from X-ray binaries (i.e., objects produced during normal star formation processes). As described in \S \ref{sec:data}, we incorporated mid- and far-IR data as well as FUV data from various sources to allow for more robust fitting of the objects' SEDs.

As discussed in Paper~I, the majority of AGN in the sample are bright enough in the mid- and/or far-IR to be observed in the \cite{Barro2019} catalog (61/72). While this suggests that AGN missed in X-ray surveys will likely be lurking in our IR sample, we do not have enough photometry to use MIR AGN diagnostics.

To very roughly estimate the remaining AGN in the sample, we note that if we separate the AGN classification by field, both GOODS-S and UDS have an AGN fraction of 13\% (in our IR-bright sample) whereas COSMOS has a much lower fraction of 4.4\%. While this is quite surprising for UDS, with a \textit{Chandra} survey exposure time of only 600 ks \citep{Kocevski2018}, the GOODS-S field features deep X-ray data that are deep enough \citep[7,000 ks;][]{Luo2017} to enable identification of the non-Compton-thick AGN. On the other hand, COSMOS has a much shallower survey depth of 160 ks \citep{Civano2016}. If we assume all fields have an AGN fraction of 13\% (for their IR-bright ELGs in particular), we would expect about 25 more AGN in our sample.

As a first-order attempt to account for Compton-thick AGN, we applied the criteria of \cite{Donley2012} and used the (observed) IRAC fluxes at $3.6$, $4.5$, $5.8$, and $8.0$~\um to identify 39 galaxies non-X-ray emitting ELGs that may contain AGN\null.  Of these, only 5 sources are in our IR sample, thus having almost no impact on our results. In this study, we treat all sources not identified through X-ray or NIR channels as normal star-forming galaxies.

To derive galaxies' physical properties (e.g., their stellar mass, star formation rate, and UV dust attenuation), we used \mcsed \citep{Bowman2020}, a flexible SED-fitting code which employs Markov Chain Monte Carlo (MCMC) Bayesian methods and allows the user to explore different star formation histories and dust attenuation laws in the fitting process.  For this study, we adopted a binned star formation history \citep[similar to that advocated by][]{Leja2017}; the \cite{Noll2009} generalization of the \cite{Calzetti2000} dust attenuation law \citep[see also][]{Kriek2013}; and the \cite{DraineLi2007} dust emission model. A full description of our methods and the various underlying assumptions used in the SED fits is given in \citet{Bowman2020} and in Paper~I.

\section{Results}\label{sec:results}

\subsection{Galaxy SED Fitting} \label{subsec:individual}
We inferred the physical properties of the CANDELS fields ELGs by fitting the galaxies' rest-frame UV through mid- and far-IR spectral energy distributions with \mcsed using the specific assumptions listed below.

$\bullet$  We build complex stellar populations (CSPs) from a linear combination of simple stellar populations (SSPs) using the Flexible Stellar Population Synthesis \citep[FSPS;][]{Conroy2009, Conroy2010SPSM} library of Padova isochrones \citep{Bertelli1994,Girardi2000, Marigo2008} and a \citep{Kroupa01} initial mass function.

$\bullet$ Nebular line and continuum emission is modeled by interpolating on a grid of \cloudy models \citep{Ferland1998,Ferland2013,Byler2017} as a function of metallicity and ionization parameter.

$\bullet$ We employ a binned star formation history with the following edges in terms of log years: $[8,8.5,9,9.5,9.8,10.12]$. The SFR within each bin is assumed to be constant.

$\bullet$ For the dust attenuation law, we use the parameterization of \cite{Noll2009} and \cite{Kriek2013}.  This generalization of the \citet{Calzetti2000} attenuation cure, which we call the ``Noll'' law, has three parameters: the total amount of attenuation, $E(B-V)$, the strength of the attenuation bump at 2175~\AA, and the difference between the wavelength dependence in the UV and that given by the Calzetti law, i.e., the ``UV slope'', $\delta$.  

$\bullet$ Emission from warm and cold dust is fit via the Spitzer-based silicate-graphite-PAH model of \citet{Draine2007} and described by \citet{Draine2011}.  The three free parameters of this law are the lower cutoff of the starlight intensity distribution ($U_{\rm min}$), the fraction $\gamma$ of dust heated by starlight with $U > U_{\rm min}$, and the PAH mass fraction ($q_{\rm PAH}$). When energy balance is not assumed, the total dust mass is a fourth free parameter and is used to normalize the dust emission spectrum. Otherwise, dust mass is derived by equating the energy attenuated by dust to that emitted in the mid- and far-IR.

$\bullet$ Based on the high average [\ion{O}{3}]/\Hb ratio of between 2.5 and 6 (see Paper~I), we fix the ionization parameter at $\log U=-2.5$. Meanwhile, we let metallicity be a free parameter. See Paper~I for more details.

$\bullet$ The fluxes measured for \Hb and \OIII are given weights equivalent to individual photometric measurements.  \Ha is given a weight equivalent to $2.5$ photometric data points. See Paper~I for a justification of these choices.

$\bullet$ For each individual galaxy, our MCMC process uses 100 walkers (random initializations) and 1,000 chains (number of steps taken by each walker) to explore the multi-dimensional parameter space.

$\bullet$ We fit the SED of each galaxy twice: once with the assumption of energy balance, and once with energy balance assumption turned off (i.e., with dust mass as a free parameter).

\begin{figure*}
    \centering
    \resizebox{\hsize}{!}{
    \includegraphics{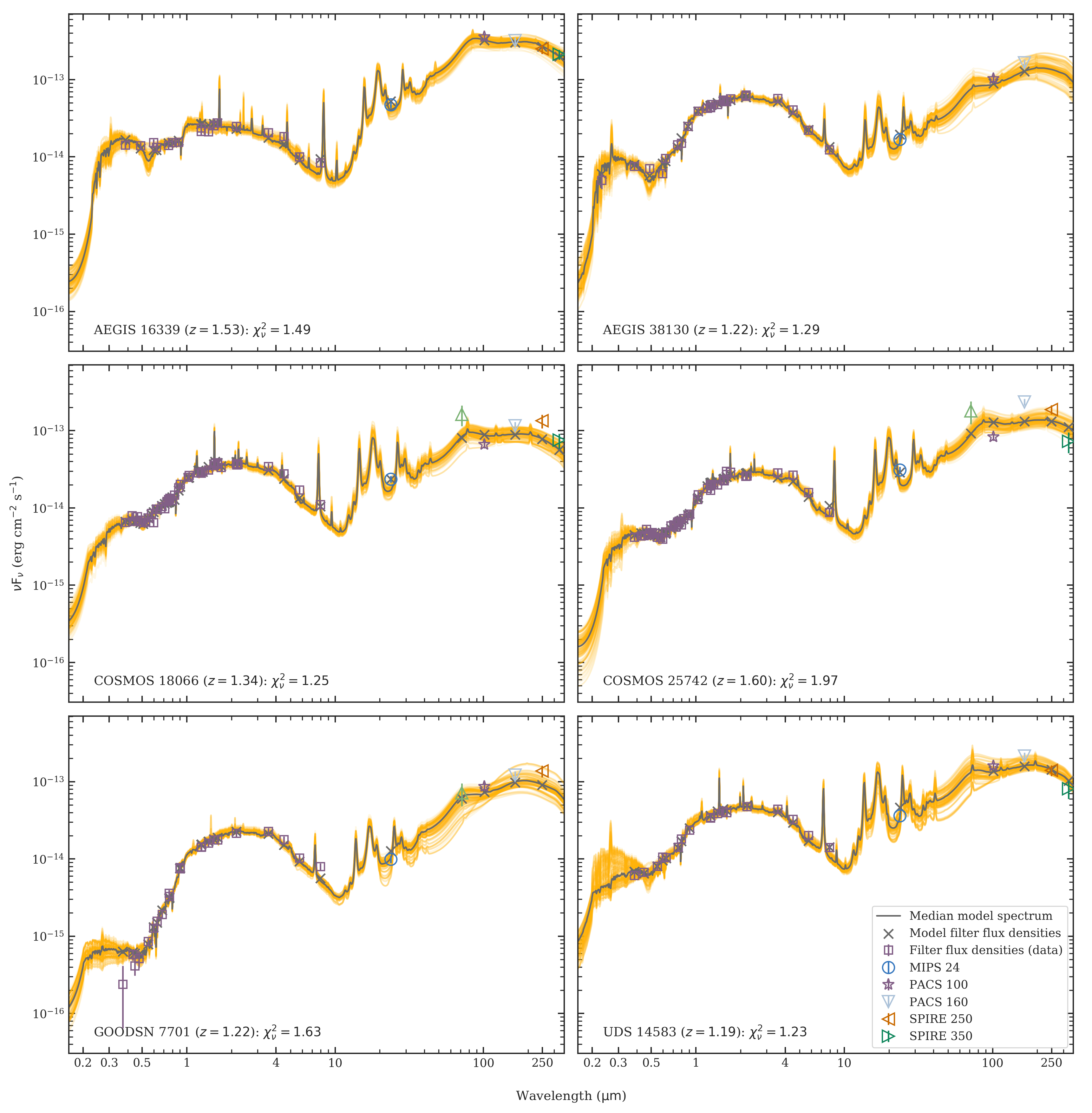}}
    \caption{Spectrum derived from two hundred \mcsed posterior samples (samples in yellow, median in gray) for sources AEGIS 16339 and 38130, COSMOS 18066 and 25742, GOODSN 7701, and UDS 14583, with a star formation rate history divided into 6 age bins, the \citet{Noll2009} dust attenuation law, and the \cite{DraineLi2007} dust model. Dust mass left and system metallicity have been left as free parameters. The observed photometric data points are shown as purple squares except for the MIR/FIR data, which have different shapes and colors depending on the instrument and wavelength. The modeled photometry is plotted as gray ``X'' marks.}
    \label{fig:samplespec}
\end{figure*}

As an example of our \mcsed fits, Figure \ref{fig:samplespec} shows the spectrum (derived from two hundred randomly chosen posterior samples) for sources AEGIS 16339 ($z=1.53$) and 38130 ($z=1.22$), COSMOS 18066 ($z=1.34$) and 25742 ($z=1.60$), GOODSN 7701 ($z=1.22$), and UDS 14583 ($z=1.19$), using the assumptions laid out above and leaving dust mass as a free parameter.  While our Bayesian framework considers both photometry and emission lines, for purposes of illustration we show just the photometry and the reduced chi-squared ($\chi_\nu^2$) value from the photometry. It is clear that in these six cases the spectra derived by \mcsed fits the data quite well.

\subsection{How Valid is Energy Balance?}\label{subsec:dmfvsaeb}

Out of the 4,350 objects in our sample, 669 have mid-IR and/or far-IR measurements.  Although this mid-IR coverage is scant (only one to five wide-band photometric measurements per galaxy), the data allow us to look for trends in the dust emission and make rudimentary conclusions about properties such as total dust mass. Moreover, our multiwavelength SED also allows us to investigate the assumption that all energy attenuated in the UV and optical is re-emitted by dust in the MIR/FIR. 

Concerning this latter point, the measured overall dust attenuation of an unresolved source depends strongly on the intrinsic star-dust geometry.  In contrast, the dust emission from this same source depends weakly, if at all, on the geometry. For example, in a highly clumped dust distribution, the light from many stars may pass through the galaxy to us with minimal attenuation, or, alternatively it may be so highly obscured as to render the galaxy invisible (and thus not appear in an emission-line selected sample). On the other hand, with a smoother dust distribution, the majority of stars will be at least somewhat shrouded in dust; in this case the galaxy would still appear in our sample, but with a larger mean attenuation.  In both cases, if the total dust mass, composition, and temperature profile are similar, the measured dust emission spectrum will be the same, as the mid- and far-IR emission suffers very little absorption or scattering. Therefore, the principle of energy balance between the sight-line-dependent attenuation and the relatively isotropic emission, may appear to be violated.

As described in \S \ref{subsec:individual}, the \mcsed SED fitting code allows users to explore how the assumption of energy balance effects our the conclusions made about galaxy properties.  When energy balance is assumed, \mcsed forces the integral over the IR spectrum to be equal to total amount of luminosity attenuated by dust in the UV/optical part of the spectrum.  This integral, of course, is directly proportional to the total dust mass, thus fixing the latter quantity. If energy balance is not assumed, then dust mass is a free parameter that scales the mass-specific \cite{DraineLi2007} models directly to the IR photometry, thus minimizing the interdependence between the UV-NIR and MIR/FIR regions of the spectrum.

To model dust attenuation, we use the dust law parameterization by \citet{Noll2009} and \citet{Kriek2013} (as described in \S \ref{subsec:individual}). The parameters of this model define the attenuation curve that is applied to the intrinsic spectrum of the complex stellar population. When energy balance between dust attenuation and emission is assumed, the MIR/FIR photometry end up influencing the attenuation curve by forcing of the attenuation and emission integrals to be the same.

As a first step towards testing the assumption of energy balance, we compare the SFRs derived by \mcsed in the case where dust mass is a free parameter (DMF) and where we assume energy balance (AEB) to those from various locally-calibrated SFR indicators \citep[see][and references therein]{Kennicutt2012}.  However, one of the most commonly used SFR indicators, the FUV stellar flux density, is somewhat compromised, in that for four of our five fields, the \textit{GALEX} NUV filter (effective wavelength $2304.7$~\AA), provides the only data point for our lower redshift galaxies. Moreover, in the fifth field (GOODS-S) where deep \textit{Swift}/UVOT data exists, just 10\% of (12/127) of galaxies with mid- and far-IR measurements have uvm2 (effective wavelength $2245.8$~\AA) and/or uvw1 (effective wavelength $2580.8$~\AA) detections. As the result, for many of our galaxies, the FUV continuum is not well constrained, and SFR indicators based the un-attenuated FUV flux density are not very reliable. For this reason, we consider the FUV only when using an IR-based correction \citep{Hao2011}.

For our analysis, we consider five SFR indicators: the NUV \citep[assumed to be centered at 2300 \AA;][]{Murphy2011,Hao2011}, \Ha \citep{Murphy2011,Hao2011}, mid-IR emission \citep[24 \um;][]{Rieke2009}, far UV flux \citep[1600 \AA;][]{Murphy2011,Hao2011}, and the total IR flux \citep[3-1100 \um;][]{Murphy2011,Hao2011}. We average the results from these five indicators, both to achieve a more complete comparison and to obtain a crude estimate of the uncertainty in the quantity.  

For FUV, NUV and \Ha, we use IR-based corrections \citep{Kennicutt2009,Hao2011} to estimate the un-attenuated luminosities, which are then converted into SFRs.\footnote{The NUV, FUV, and MIR luminosities are calculated from modeled SED fluxes rather than photometric measurements, as our wide-range of redshifts makes associating individual filters with rest-wavelengths difficult.}The IR-based attenuation corrections are derived from energy balance arguments; an example of such a relationship is given by \citet{Hao2011} who used the IRX-$\beta$ relation (see discussion later in this section) and NUV-FUV color to determine IR corrections for their SFR indicators.

\begin{figure*}[!ht]
    \centering
    \resizebox{\hsize}{!}{
    \includegraphics[scale=0.61]{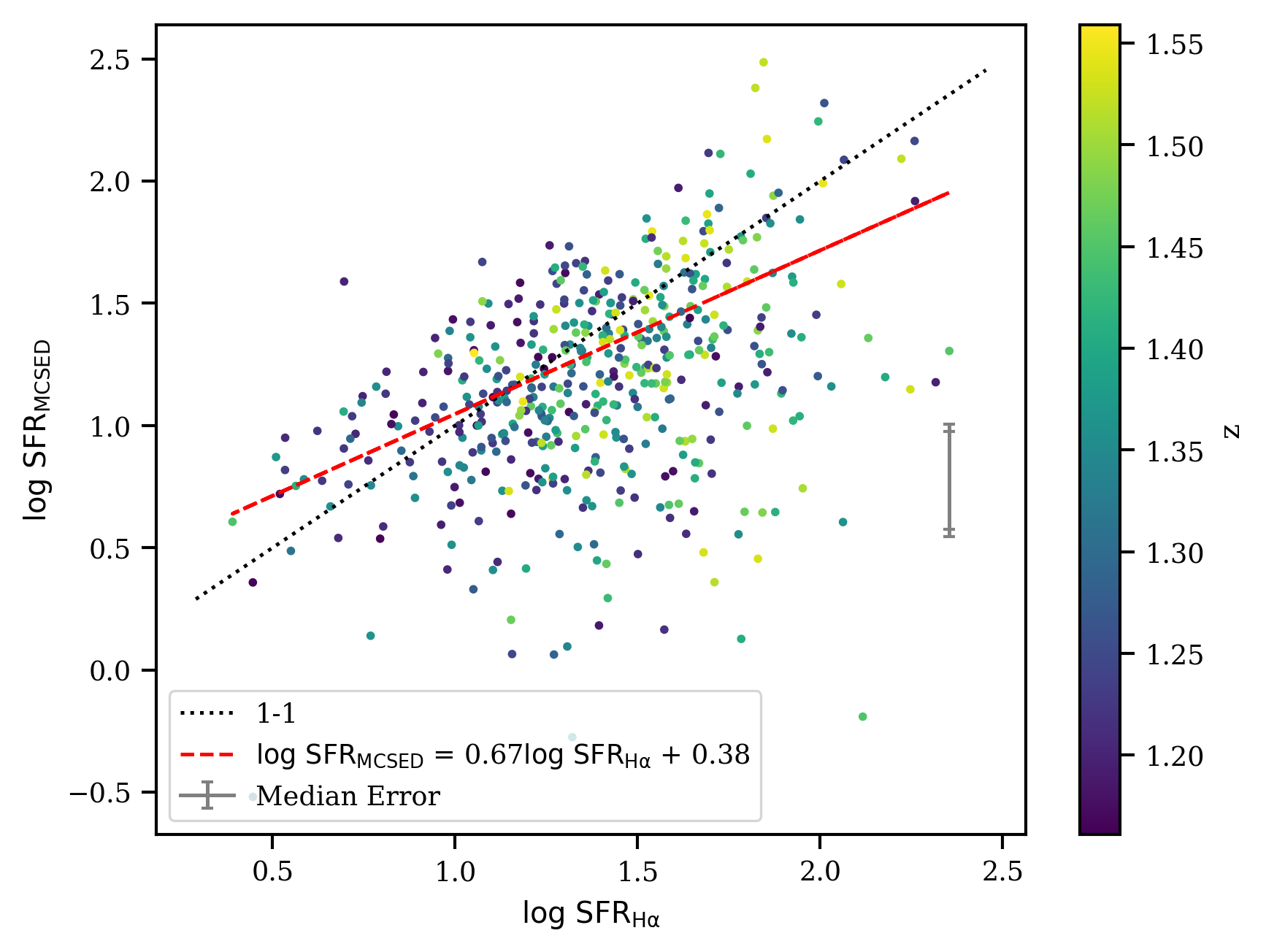}
    \includegraphics[scale=0.61]{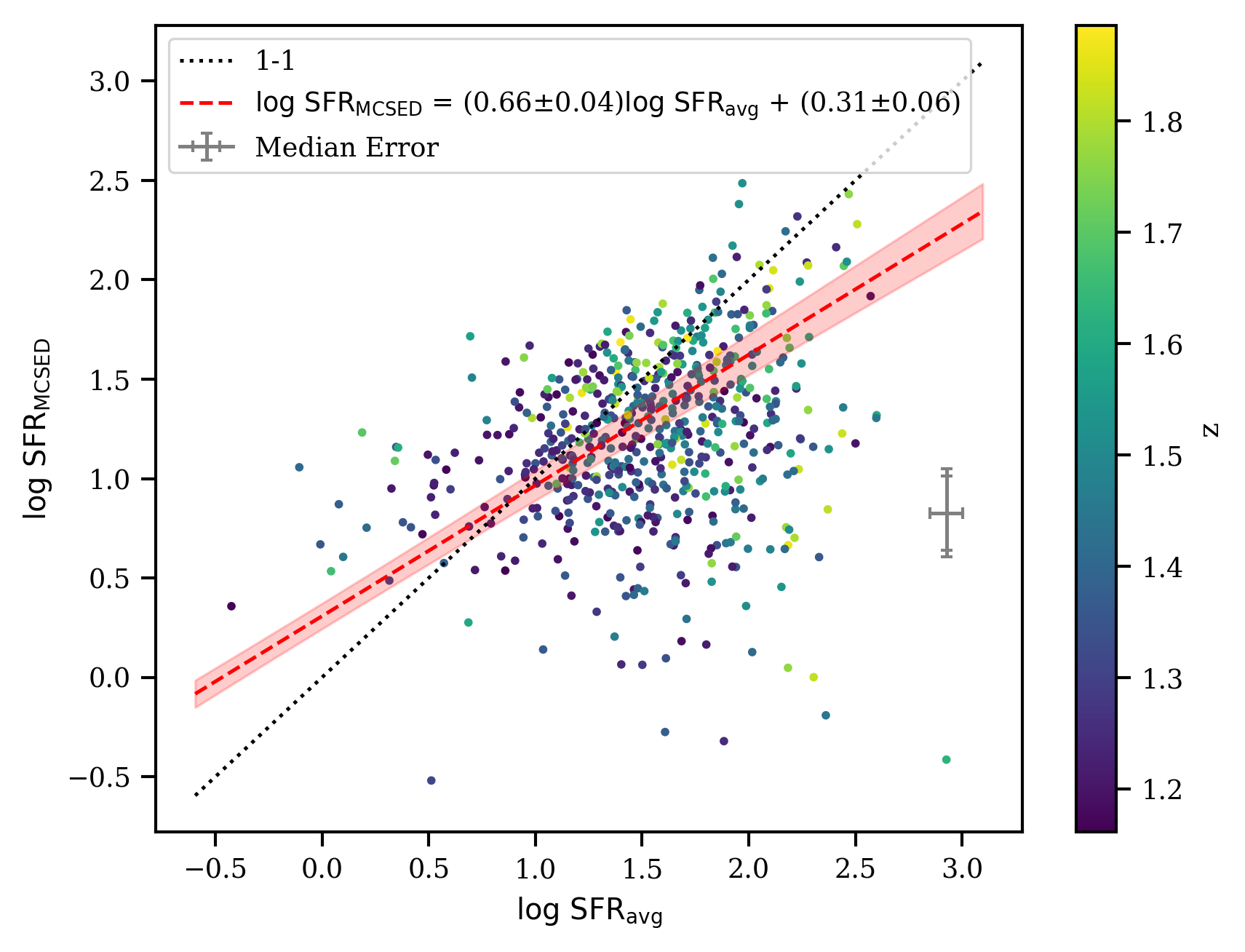}}
    \resizebox{\hsize}{!}{
    \includegraphics[scale=0.61]{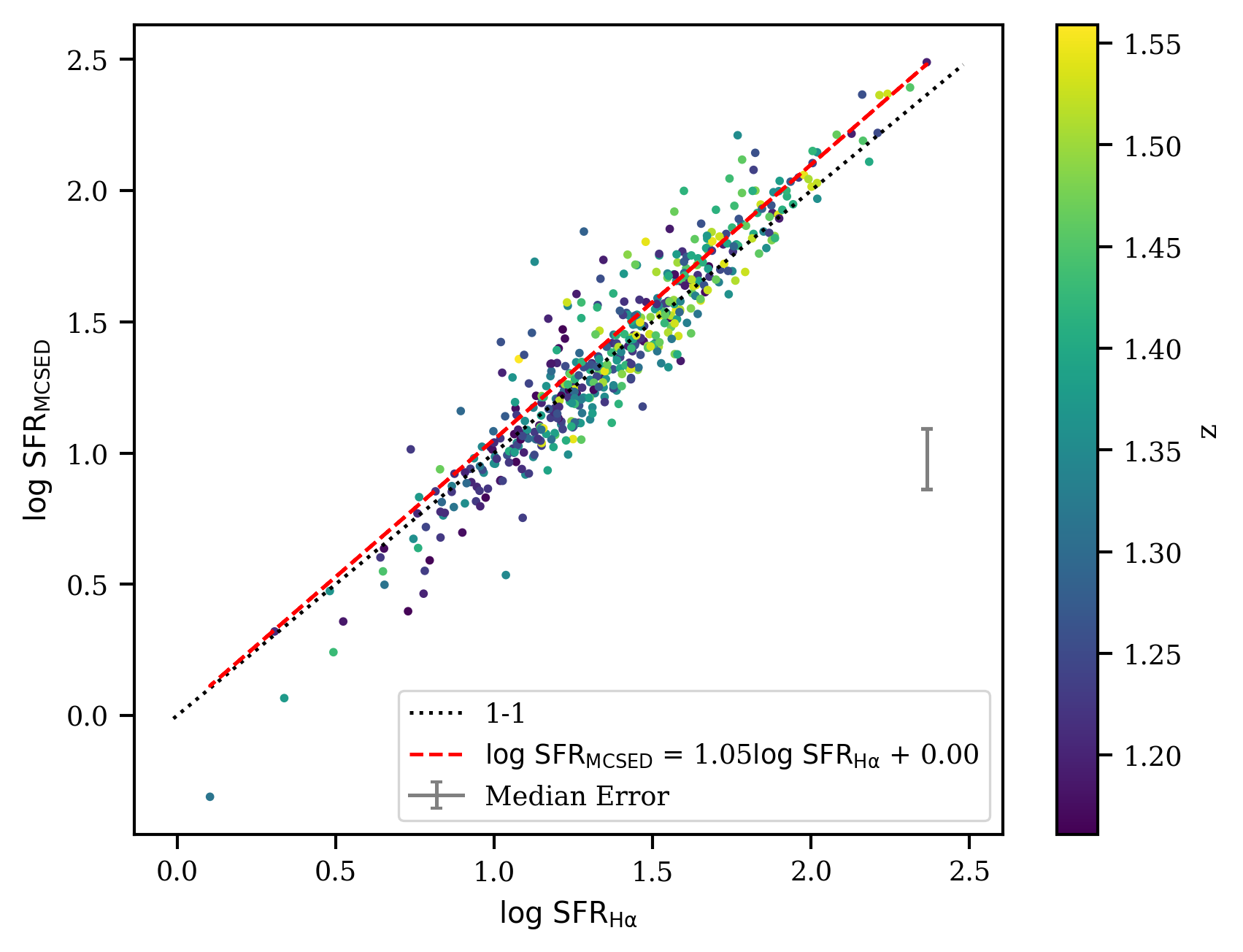}
    \includegraphics[scale=0.61]{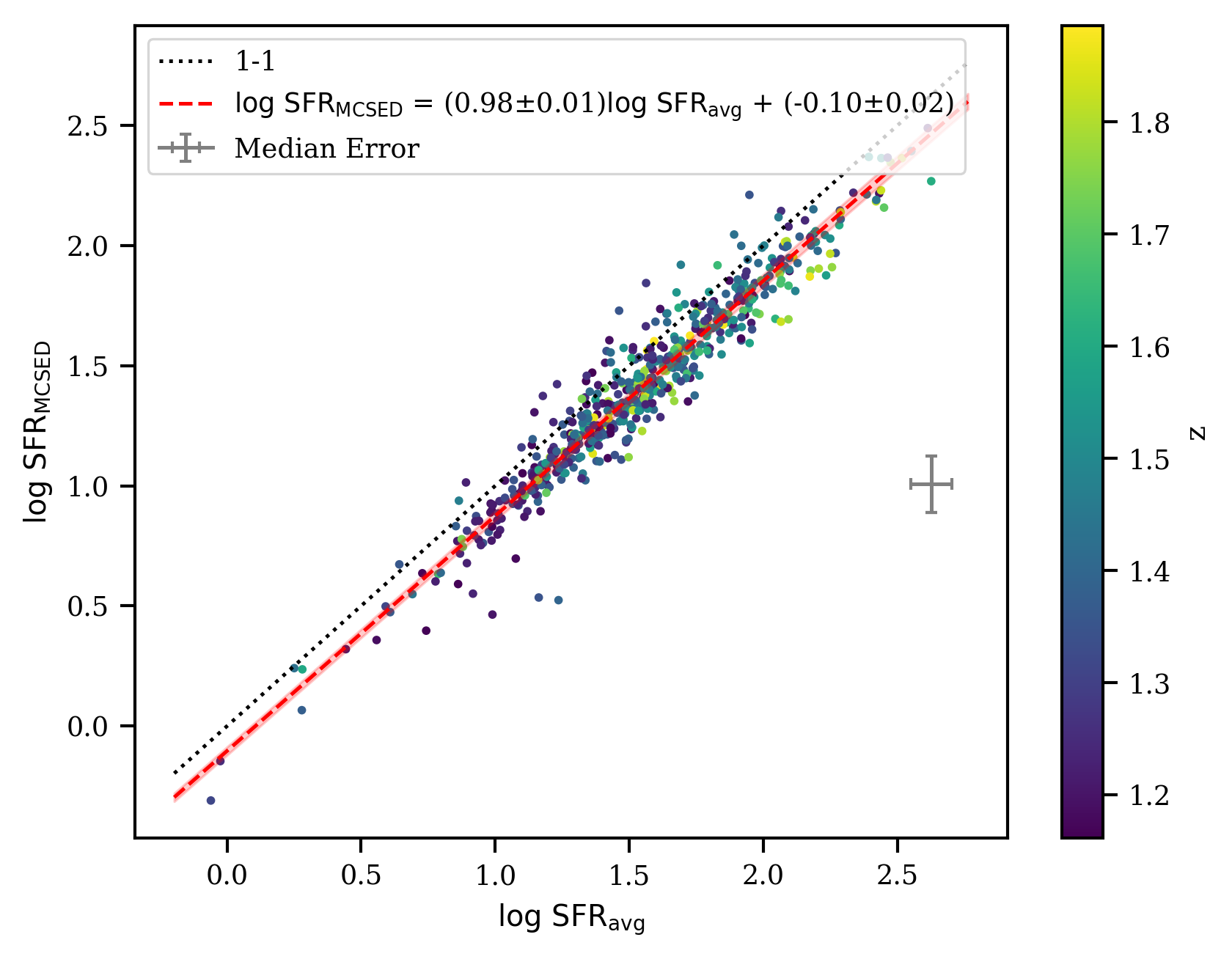}}
    \caption{Comparison of \mcsed SFRs (modeled to be constant over the last 100 Myr) with H$\alpha$-based SFRs (left) and the average of the locally
    calibrated SFR indicators (right).  
    In the top two plots, the \mcsed fits treat dust mass as a free parameter; in the lower plots, energy balance between dust emission and absorption is used to help determine the dust mass. We show the best-fit line taking \mcsed SFR errors into account (dashed red line) and the one-to-one relation (black dotted line).}
    \label{fig:sfrcomp}
\end{figure*}

To these wavelength-specific SFR indicators, we also add the SFR determined by \mcsed over the past $10^8$~years, i.e., the size of the most recent bin in our quantized star formation history. The comparisons between the average indicator-derived SFRs, the \Ha SFRs, and \mcsed's SED-based SFRs are shown in Figure \ref{fig:sfrcomp} for both DMF and AEB fits. In each plot, we show both the best-fit line and the one-to-one relation. In the case of \Ha, the linear regression is done under the assumption of perfect indicators (given the difficulty in determining the true uncertainty in the measurements). For the average case, we take the standard deviation in the wavelength-specific SFR indicators to be a measure of the error and use orthogonal distance regression. In all cases, the points are colored by redshift.

\begin{figure}
    \centering
    \resizebox{\hsize}{!}{
    \includegraphics{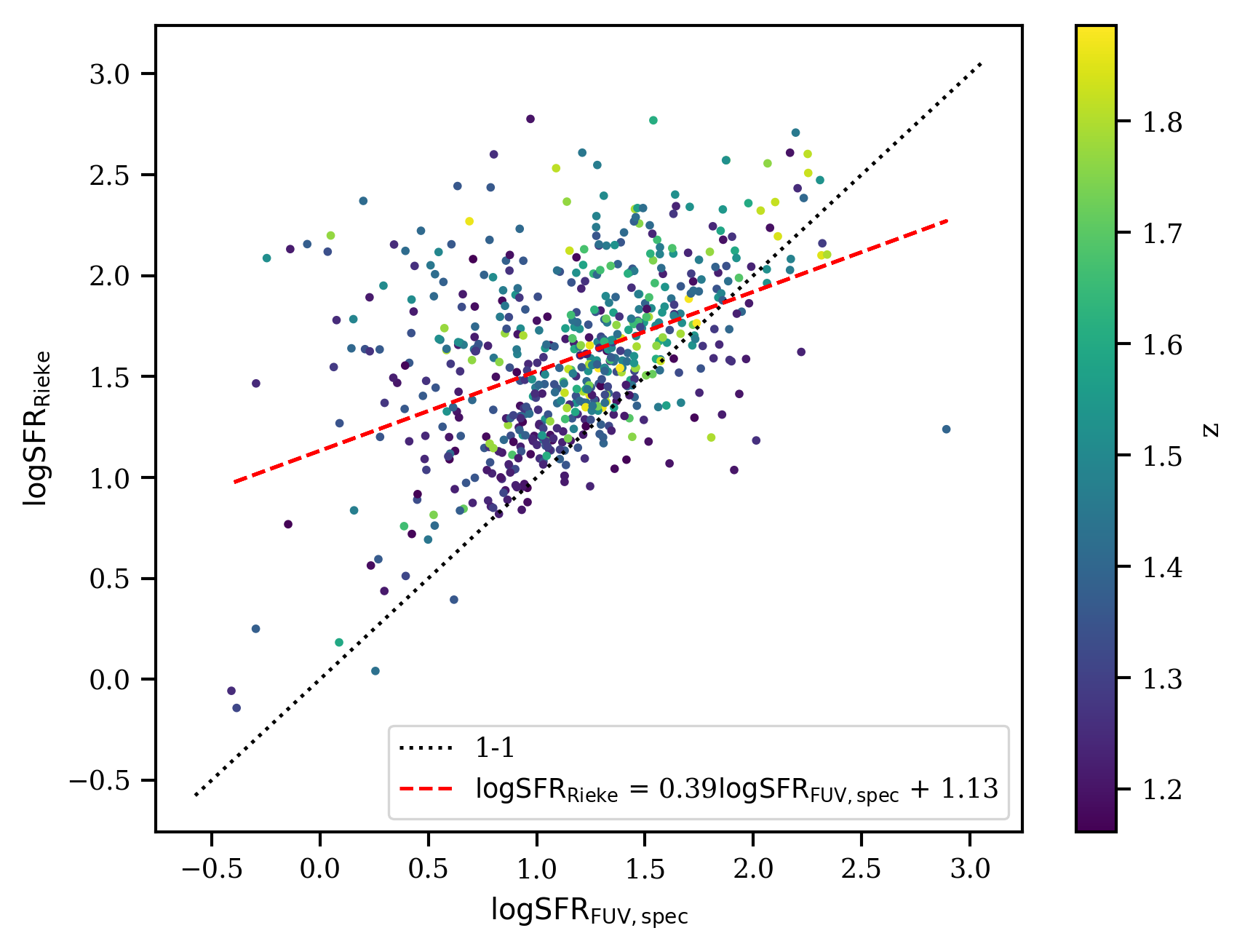}}
    \caption{Comparison of SFRs derived using the \cite{Rieke2009} 24 \um indicator and the FUV indicator \citep{Murphy2011,Hao2011,Kennicutt2012}. Un-attenuated UV fluxes have been determined using the prescription from \cite{Calzetti2001}. The disagreement is enormous and likely stems from sample biases discussed in the text. }
    \label{fig:RiekeUV}
\end{figure}

An examination of the figure shows that there are no significant trends with redshift, suggesting a lack of selection bias in that regard. However, a more nefarious issue is that of IR-selection effects. Our sample of galaxies is naturally biased toward systems with either 1) relatively low dust attenuation given stellar mass (driven by our original emission-line selection criteria) or 2) strong dust emission (necessary for detection in the mid- and far-IR). These biases are clearly seen in Figure~\ref{fig:RiekeUV}, which compares the \cite{Rieke2009} 24~\um and FUV SFR indicators with $\beta$, assuming  $A_{1600}=2.31(\beta + 2.35)$ \citep{Calzetti1994,Calzetti2001,Bowman2019} and $\log {\rm SFR_{UV}} = \log {\rm L_{1600}} - 43.35 \msyr$ \citep{Murphy2011,Hao2011,Kennicutt2012}.  As mentioned above our FUV flux densities are not always well-determined; nevertheless, there is value in isolating the UV SED from the IR.

In Figure \ref{fig:sfrcomp}, we can see the drastic difference between letting dust mass be a free parameter and assuming energy balance. In the former case, the scatter in the correlation dominates and the points do not seem to be consistent with a one-to-one relation. In the latter case, the scatter is quite limited and for the most part, the \mcsed SFRs agree with those derived from the wavelength-specific SFR indicators, with only a small offset. 

However, this finding does not necessarily imply that the assumption of energy balance leads to more accurate predictions. The concept of energy balance is encoded into the SFR indicators themselves, as the IR corrections required to infer un-attenuated FUV, NUV, and \Ha SFRs use this assumption, as do the \cite{Rieke2009} MIR and \cite{Hao2011,Murphy2011} total IR indicators. Therefore, it is quite possible that adhering to the energy balance argument simply makes the output of \mcsed similar to that of the other indicators.

\begin{figure*}
    \centering
    \includegraphics{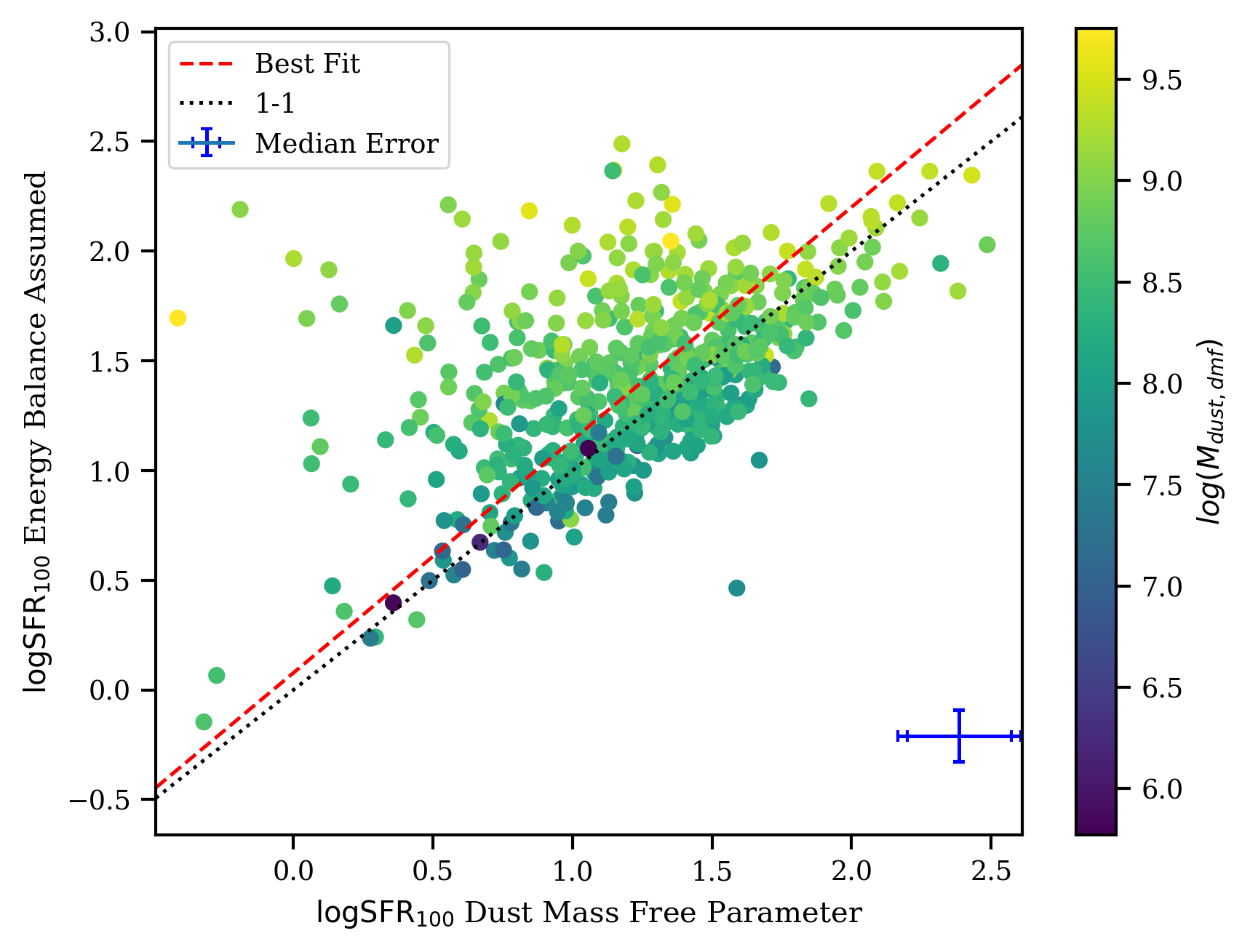}
    \caption{Comparison of \mcsed SFRs when dust mass is a free parameter (DMF, $x$-axis) to SFRs when energy balance is assumed (AEB, $y$-axis). The red dashed line is the best-fit orthogonal distance regression; dotted black line shows the one-to-one relation. The points are more consistent with a one-to-one relationship, but the scatter can largely be explained by the DMF dust masses, which are a proxy for the MIR/FIR flux measurements.}
    \label{fig:dmfvsaeb}
\end{figure*}

We explore this issue further in Figure \ref{fig:dmfvsaeb}, where we directly compare the SFRs derived by \mcsed when dust mass is free (DMF) with those found when energy balance is assumed (AEB). The colors of the points represent the galaxies' dust masses. The figure demonstrates that the majority of galaxies scatter around the best-fit line (which is not too different from a one-to-one relationship) and this scatter can be explained by the DMF dust masses, which serve as a proxy for the galaxies' MIR/FIR flux densities.  In other words, dust mass is the normalization that forces a match to the absolute flux measurements in the mid- and far-IR\null.  Consequently, higher MIR/FIR fluxes tend to increase the AEB SFRs more than the DMF SFRs: the DMF SFRs are less dependent on the IR SED.

An important implication of this phenomenon is that any confusion noise in the IR data would have a greater impact on SFR estimates when energy balance is assumed. Indeed, some of the disparity between the \cite{Rieke2009} 24-micron and the FUV SFR indicators in Figure \ref{fig:RiekeUV}, in which energy balance is assumed, could stem from confusion noise in addition to the biases discussed earlier.

\begin{figure*}
    \centering
    \includegraphics{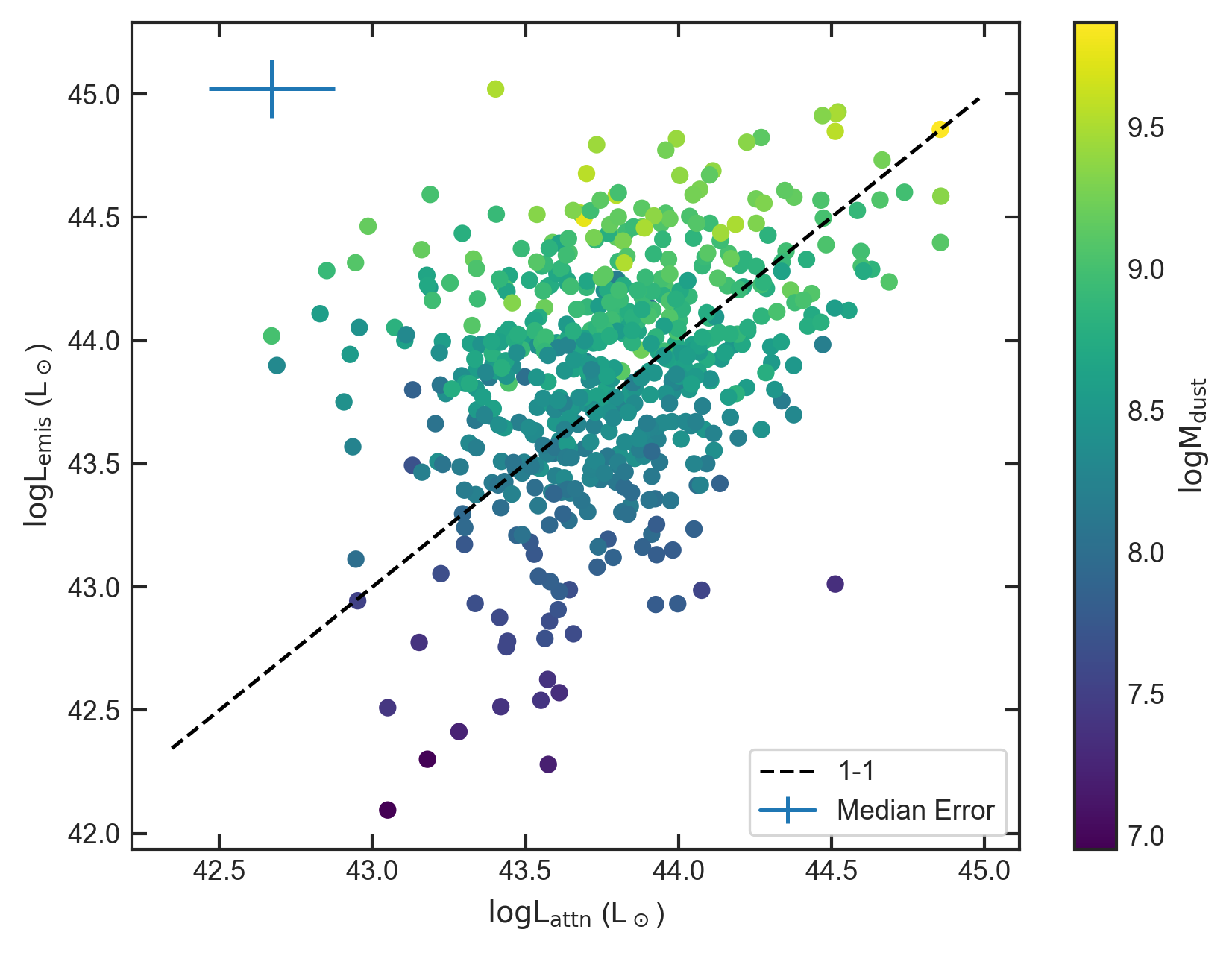}
    \caption{Dust emission vs.~attenuated luminosity in the case where energy balance is not assumed. The points are colored by the dust mass measurements. While the one-to-one line serves as a good ensemble average, the scatter is large ($\sim 0.54$ dex). Differences in dust mass contribute most to the scatter, which could indicate a mixture of confusion noise and  physical variations in galaxies with different dust contents.}
    \label{fig:emisvsattn}
\end{figure*}

As the culmination of our evaluation of the energy balance argument, Figure \ref{fig:emisvsattn} presents a direct comparison of the dust attenuated vs.~emitted luminosity in the case where dust mass is a free parameter. While there is an overall agreement between the attenuated and emitted energy, the scatter is quite large ($\sim 0.54$ dex) and can be largely attributed to differences in the calculated dust mass, which, as mentioned earlier, is nearly a proxy for mid- and far-IR flux density, as the dust mass provides the normalization for the long wavelength SED. 

The dependence of dust mass on the attenuated vs.~emitted energies probably reflects both issues of confusion noise, as discussed earlier in reference to Figure \ref{fig:dmfvsaeb}, and real, physical differences between galaxies. For example, if dustier galaxies have their dust more uniformly distributed than their lower-dust counterparts, a result like Figure \ref{fig:emisvsattn} could arise. 

However, given the dearth of rest-frame mid- and far-IR and FUV measurements, we refrain from making stronger or more specific conclusions from these results. Furthermore, from the reduced $\chi^2$ values of the fits, it is difficult to choose between our DMF or AEB models: neither provides an objectively better fit to the data.  Of course, the mid- and far-IR data for these $1.2 < z < 1.9$ galaxies are rather sparse compared to what is available in the UV-NIR part of the spectrum.  More/better data at long wavelengths would greatly assist in discriminating between the two classes of models.

Finally, we present the IRX-$\beta$ curve for our sample, where $\beta$ is the average slope in the rest-frame spectrum between 1300-2600 \AA~and IRX is the ratio of IR to UV flux density. First found to apply to local starburst galaxies by \cite{Calzetti1997} and \cite{Meurer1999}, the relation between $\beta$ and IRX can be thought of as an expression of energy balance. In a strongly star-forming population, the rest-frame UV spectrum reflects the Rayleigh-Jeans tail of the blackbody function, which is almost a power law with index around $-2.35$ \citep{Calzetti2001}. Differences between the measured value of $\beta$ and this theoretical value arise from dust absorption, which is stronger at shorter wavelengths (i.e., reddening). Meanwhile, the majority of IR emission stems from the re-radiation of this light by dust. 

\begin{figure*}
    \centering
    \resizebox{\hsize}{!}{
    \includegraphics{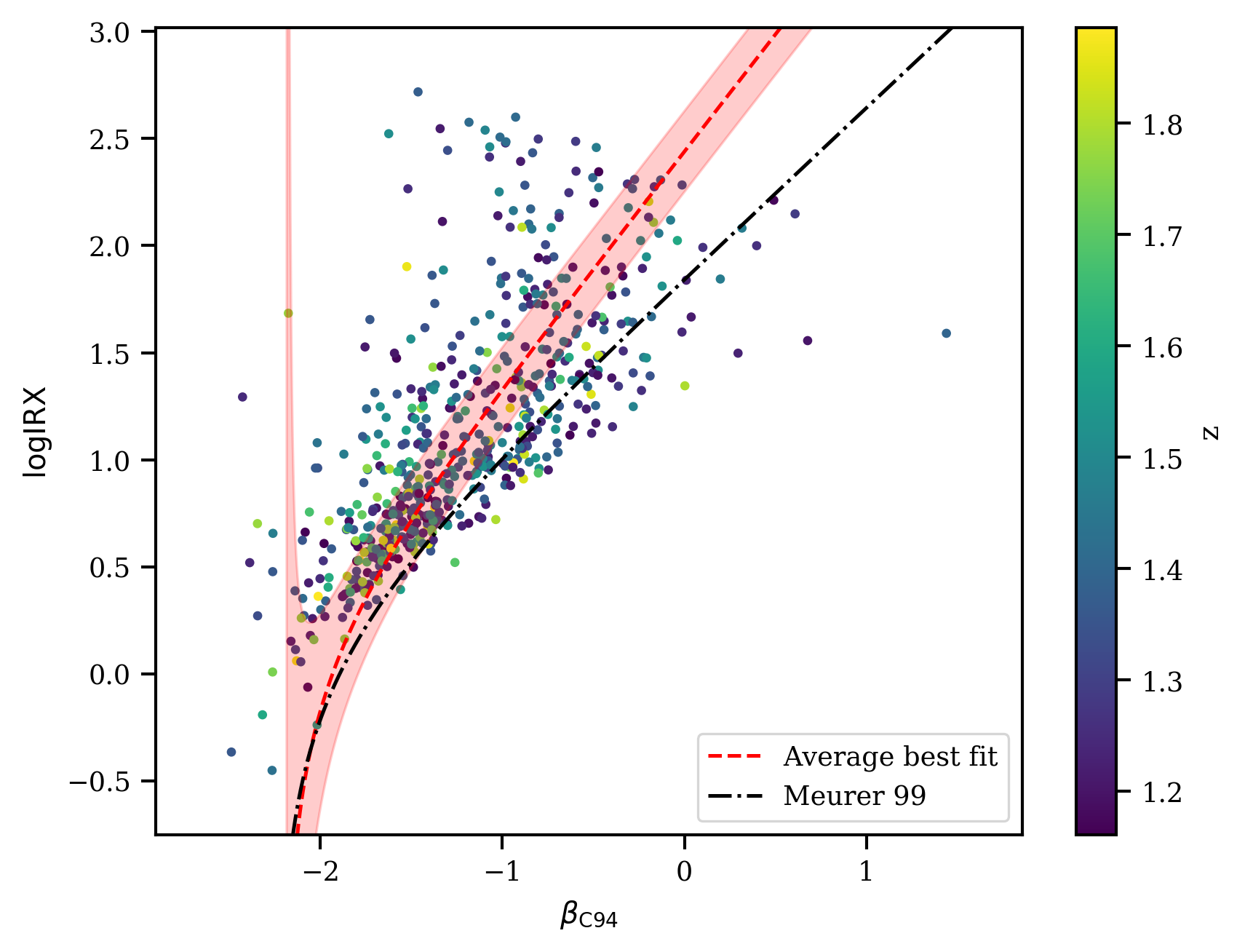}
    \includegraphics{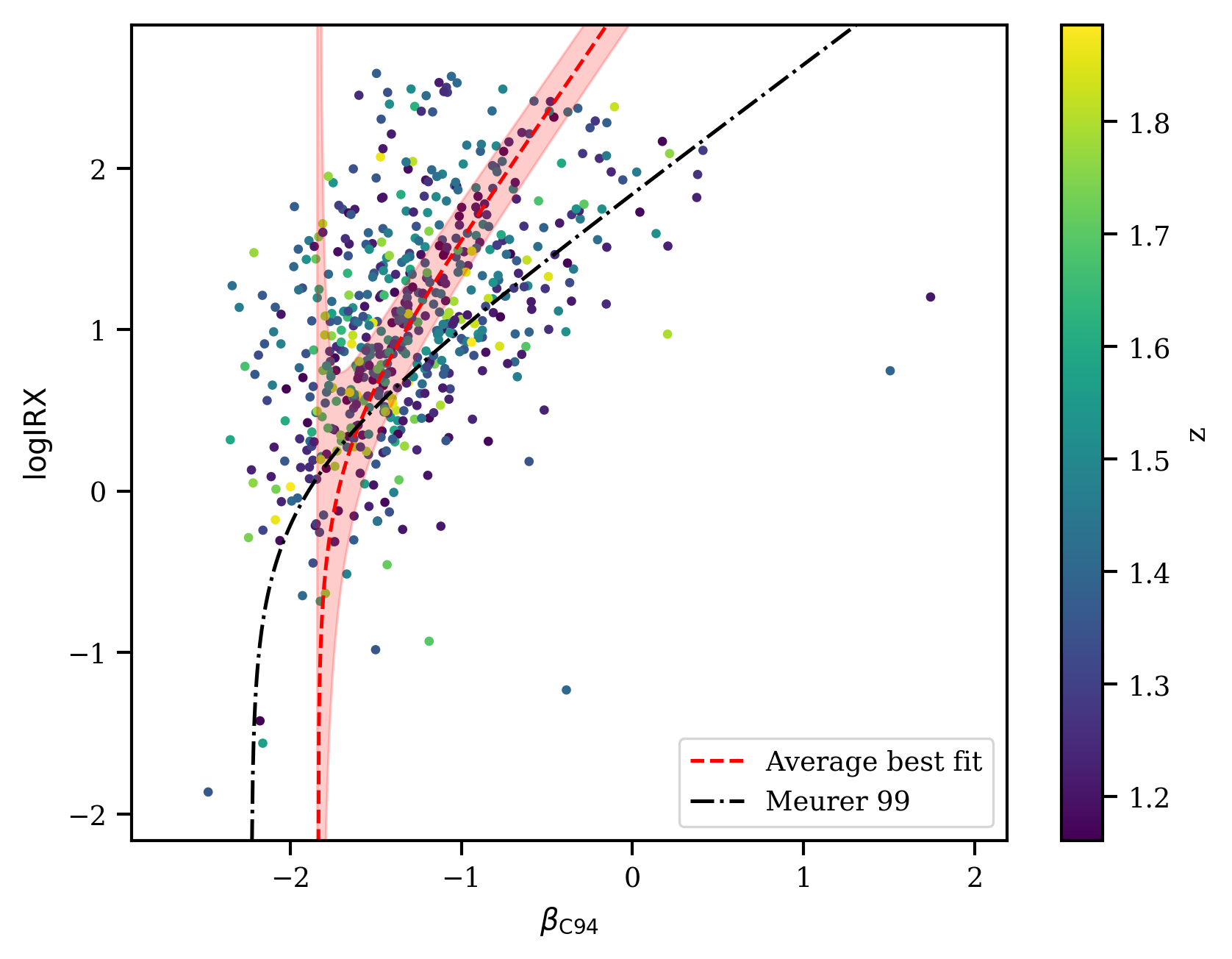}}
    \caption{IRX vs.~$\beta$ for our ELGs with long-wavelength measurements along with a best-fit function (dashed red curve with translucent pink shaded area for uncertainties) and the \cite{Meurer1999} relation (dash-dotted black curve). In the left panel, energy balance is assumed; in the right panel, dust mass is a free parameter. In both cases, the points tend to lie above the \cite{Meurer1999} relationship for larger $\beta$ values, suggesting complex geometries or low UV optical depths \citep{Narayanan2018}. In the energy balance case, all points tend to lie above \cite{Meurer1999}, implying steeper attenuation laws as well \citep{Narayanan2018}. However, Figure \ref{fig:dustnodust} suggests that the steeper attenuation laws are artificially produced from overestimated IR fluxes.}
    \label{fig:irxbeta}
\end{figure*}

However, there is usually significant scatter in the relationship, caused by variations in dust attenuation law, the presence or absence of old stellar populations, and complications associated with geometry and clumpiness \citep{Narayanan2018}.  Figure \ref{fig:irxbeta} shows our IRX-$\beta$ relation, calculated using \mcsed's best-fit spectral distribution with and without the assumption of energy balance. 

We note that given our lack of photometric data in the FUV, our $\beta$ values are not particularly well constrained; therefore, any scientific inferences we make using the IRX-$\beta$ curve require follow-up studies for confirmation.  The figure shows IRX vs.~$\beta$ along with the \cite{Meurer1999} relation and a best-fit curve using the equations of \cite{Meurer1999}. For consistency with the aforementioned study, we define $\beta$ by fitting a linear function to the rest-frame \mcsed spectrum in 10 intervals between $1268$ and $2580$~\AA, as defined by \cite{Calzetti1994}. 

Meanwhile, IRX is defined as the total FIR luminosity divided by the monochromatic luminosity at $1600$ \AA\null. The equation to derive the FIR luminosity from our SEDs is provided in the Appendix of \cite{Helou1988}:

\begin{align}
    \log \left({\rm IRX}_{1600} \right) & =  \log \left(10^{0.4A_{1600}} - 1 \right) + \log B \\
    A_{1600} & = a + b\beta
\end{align}
With the parameters $a$, $b$, and $\log B$ in the equations above, we fit IRX vs.~$\beta$ and $\beta$ vs.~IRX and average the resulting parametric fits. The best-fit values for the energy balance case are $a = 6.0 \pm 0.4$, $b = 3.8 \pm 0.2$, and $\log B = 0.0 \pm 0.1$. The values for the dust mass free case are $a = 7.1 \pm 0.5$, $b = 3.9 \pm 0.3$, and $\log B = 0.3 \pm 0.1$. 

As we see in Figure \ref{fig:irxbeta}, our sample of $1.2 < z < 1.9$ ELGs with long-wavelength detections tend to lie above the \cite{Meurer1999} relation at larger $\beta$ values for both cases. This suggests complex dust geometries and/or lower UV optical depths \citep{Narayanan2018} than those found in local starburst galaxies for IR-bright ELGs. In the energy balance case, the sample tends to lie above the \cite{Meurer1999} relation even for low $\beta$ values, suggesting steeper attenuation laws than what \cite{Meurer1999} measured for local starbursts. 

\begin{figure*}
    \centering
    \resizebox{\hsize}{!}{
    \includegraphics{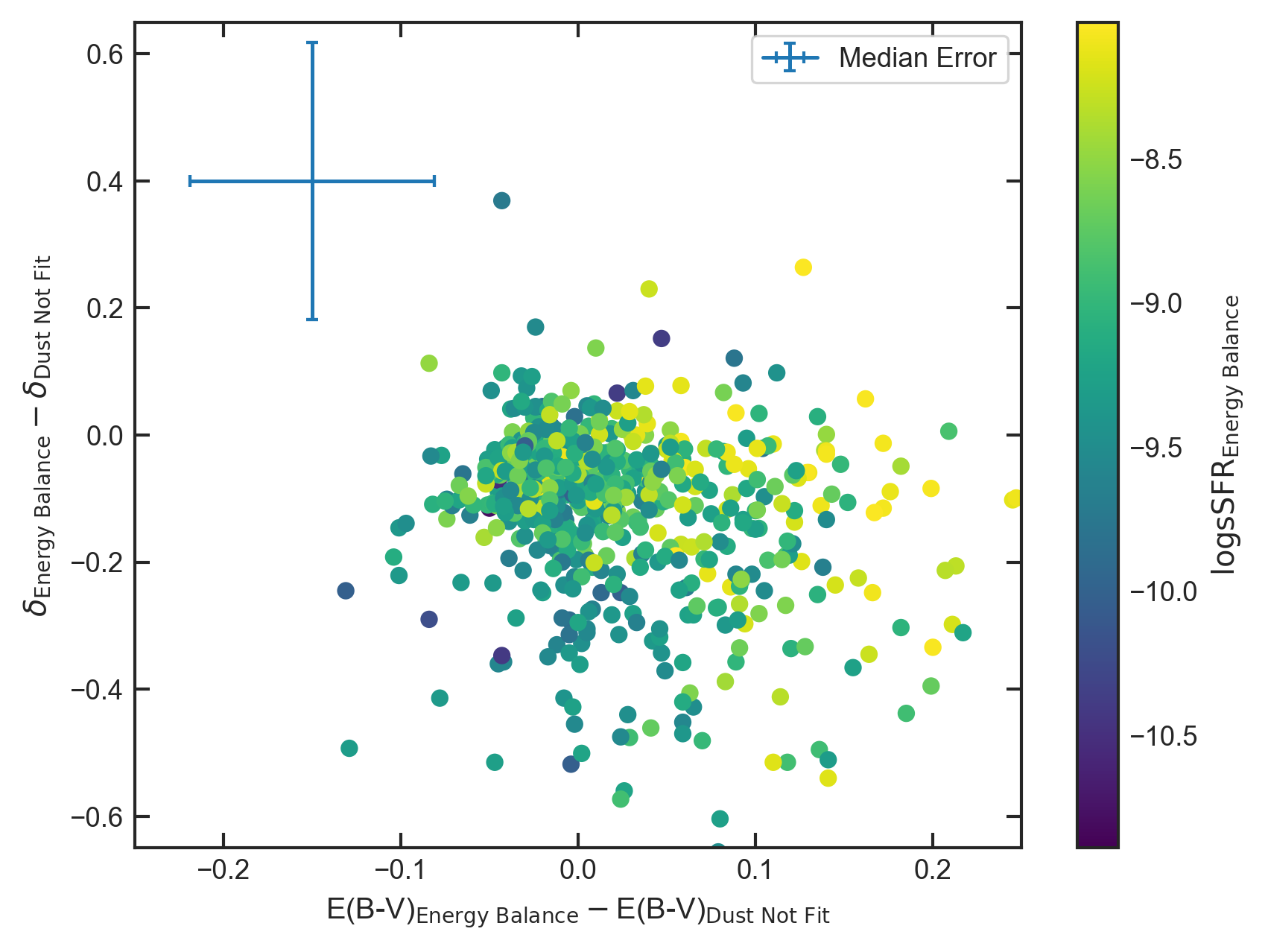}
    \includegraphics{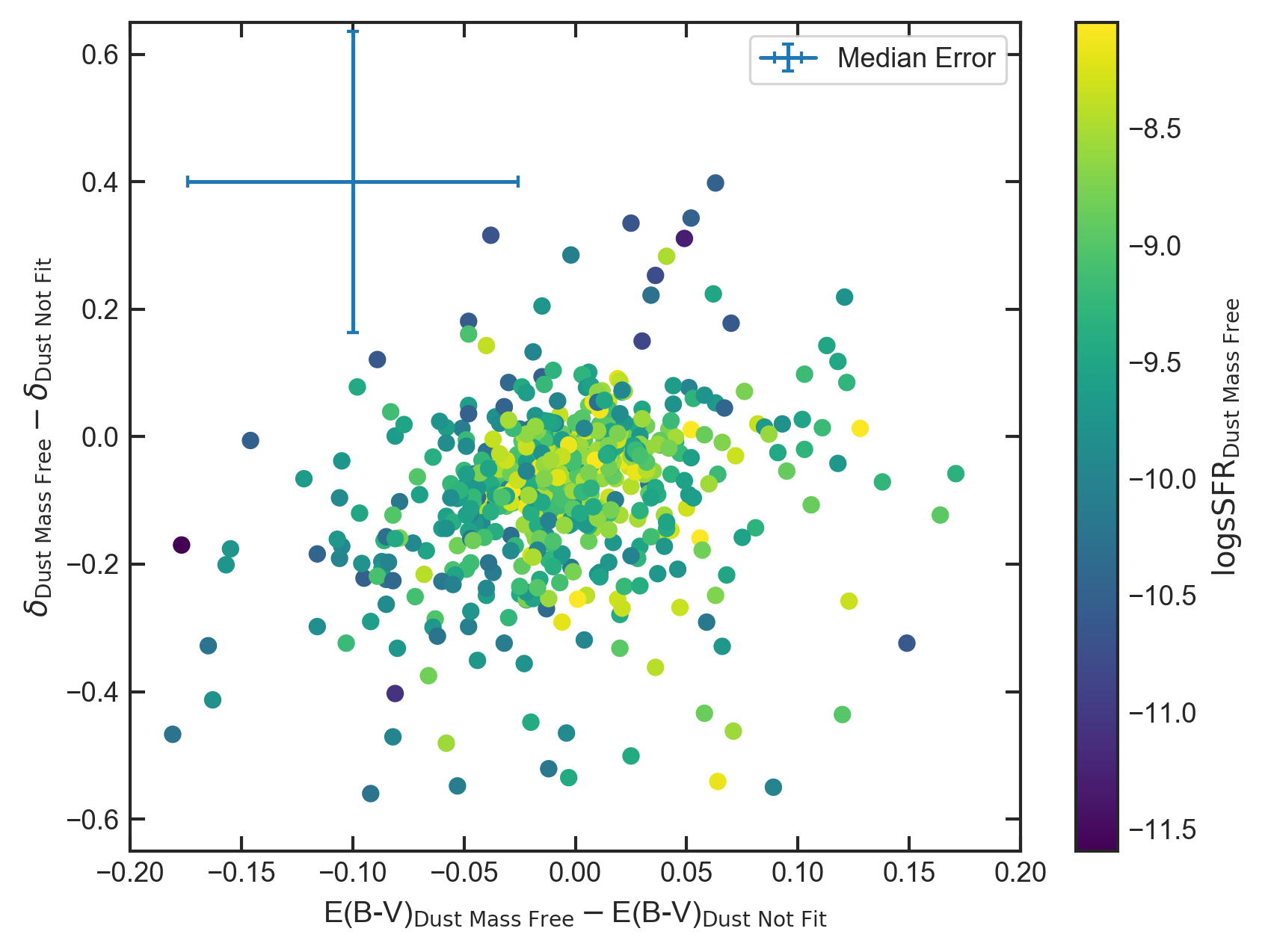}}
    \caption{Comparison of dust attenuation law parameters $\delta$ (related to UV slope) and E(B-V) for our IR sample when the IR spectrum is fit vs.~ not fit (i.e., ignored). On the left, energy balance assumed (vs.~ dust not fit); on the right, dust mass is treated as free parameter (vs.~dust not fit). When energy balance is assumed, there is a systematic shift compared to not fitting dust emission: the models suggest larger differential extinction and steeper attenuation laws, most likely stemming from overestimated fluxes caused by confusion noise in the far-IR. There is much less of a systematic shift for the dust mass free case (with somewhat steeper attenuation laws when dust is fit), which makes sense given the relative independence in treating dust attenuation and emission.}
    \label{fig:dustnodust}
\end{figure*}

However, when we compare our fits for the attenuation law using the same IR sample with and without fitting the IR spectrum (Figure \ref{fig:dustnodust}), we find that fits which include the mid- and far-IR and require energy balance yield systematically larger differential extinctions (E(B-V) larger by 0.03 on average) and steeper attenuation laws ($\delta$ lower by 0.13 on average) than fits that do not include the IR.  While there is considerable scatter when dust mass is unconstrained, we find no difference in E(B-V) and a smaller preference toward steeper attenuation laws ($\delta$ smaller by 0.09 on average) when fitting the IR spectrum. In general, we can see that fitting the IR spectrum affects the dust attenuation parameters much more strongly when energy balance is assumed, which makes sense considering the previous discussion.

Given the information in Figures \ref{fig:RiekeUV} and \ref{fig:dustnodust}, we suspect that confusion noise in the mid- and far-IR is causing some fluxes to be overestimated, and this impacts the UV SED fitting when energy balance is assumed. Of course, additional rest-frame FUV and improved spatial resolution in the mid- and far-IR  would allow us to calculate much more reliable $\beta$ and IRX values.

\subsection{Relation among sSFR, Stellar Mass, and \mdust} \label{subsec:FP}

Using the \mcsed-estimated SFR and stellar masses, we introduce sSFRs with the definition $\log {\rm{sSFR}} = \log {\rm{SFR}} - \log M_*$. While sSFR has clearly non-zero covariance with stellar mass given its definition, the fact that it is an intensive rather than extensive property like SFR and stellar mass makes it an interesting quantity to estimate and analyze.

Under the assumption of energy balance, one of the most prominent features we observe is the connection between sSFR, stellar mass ($M_*$), and dust mass. When energy balance is not assumed, a different relation exists among these variables, albeit with considerably more scatter. In this section, we delve into the relationship between these three variables with and without the energy balance constraint. An important point to note is that when energy balance is assumed, the effective covariance between the parameters is increased due to the effect that the mid- and far-IR has on solutions for the UV-NIR spectrum. Therefore, even though we present the relationship with energy balance, we avoid drawing strong conclusions from it.


\begin{figure*}
    \centering
    \includegraphics{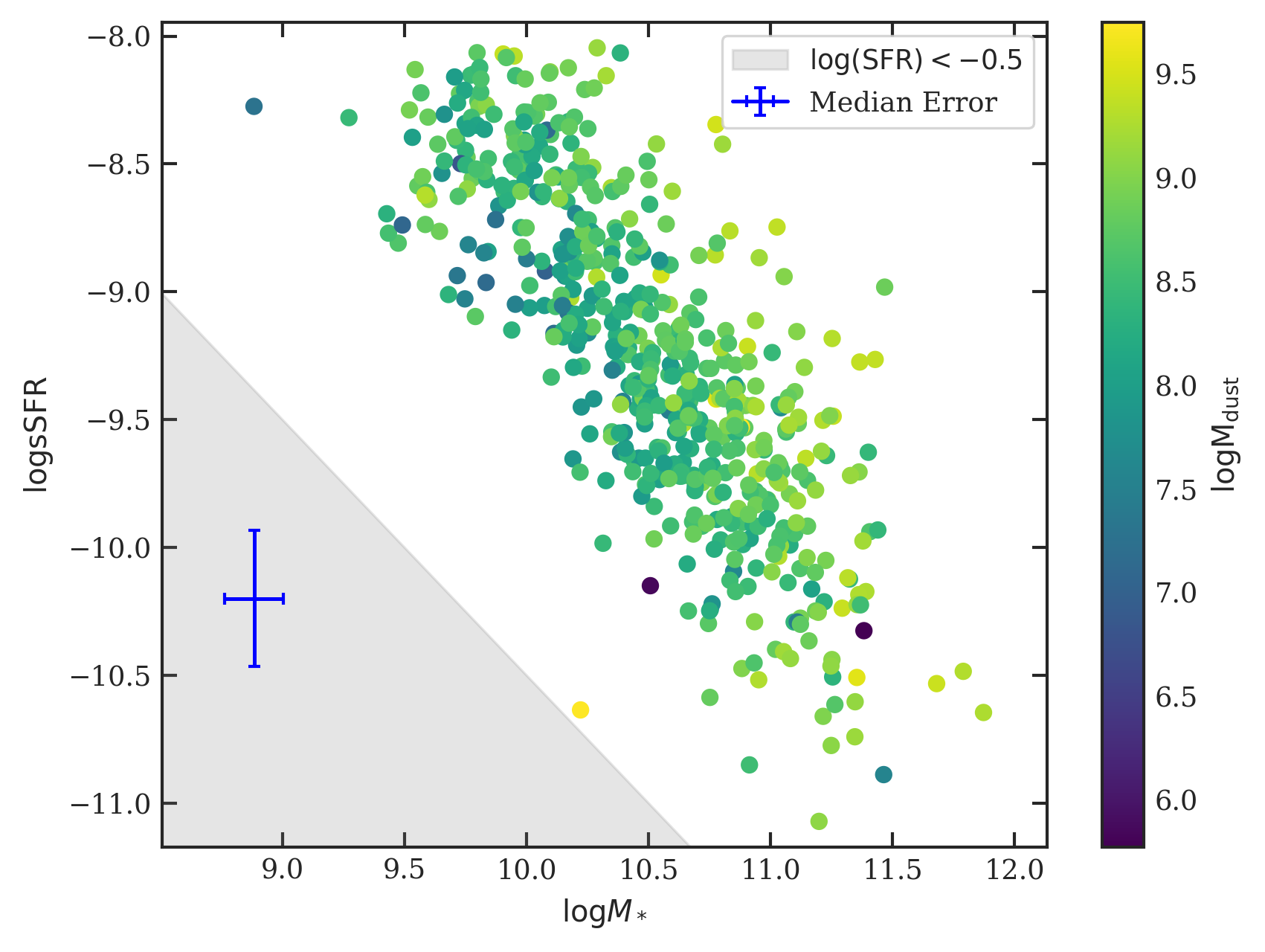}
    \caption{The sSFR vs.~$\log M_*$ relation colored by $\log$~\mdust (when dust mass is left as a free parameter). Areas of the parameter space with zero completeness are shown in gray. While incompleteness sets in well before the gray area and probably plays a role in shaping our observations, there is likely a relationship among the three variables, albeit with sizable scatter.}
    \label{fig:logMssFRdust}
\end{figure*}

Figure \ref{fig:logMssFRdust} shows the $\log M_*$ vs\ sSFR relation colored by $\log M_{\rm dust}$ (when dust mass is a free parameter). There is a clear correlation between the two quantities, with \mdust appearing partially responsible for the scatter, as it tends to change in a direction almost orthogonal to the principal correlation. Put another way, if we fix the dust mass, the correlation between stellar mass and sSFR is much tighter. 

Also shown in the plot are the regions of parameter space where objects may be lost due to selection effects. Given our objects were selected both for the presence of emission lines and their detectability in the mid- and far-IR, our principal constraint is in SFR, as below certain values the completeness in emission lines and IR photometry decreases. In addition, there is a constraint in stellar mass, as our objects must have apparent magnitudes of $m_{JH}<26$ for a slitless spectral extraction. Furthermore, from Figure \ref{fig:sampleprop}, we observe that while there is no hard mass cutoff, the vast majority of sources with mid- and far-IR data have significantly higher masses than for the parent ELG sample.

For simplicity, we define the borderline SFR as roughly where such objects are not found in our sample. Incompleteness probably sets in around $\sim 0.5$ dex above the border line (in SFR), which is close to the locus of the data, suggesting that incompleteness does play a role in shaping the relation we see. Nevertheless, given the trend in the region where observations are complete, the relationship between the three variables is likely to be real, although the scatter is certainly significant.

We created a linear model relating stellar mass to sSFR and \mdust, ignoring the heterogeneous uncertainties in the parameters. When we assume energy balance, we find the best-fit equation is
\begin{multline} \label{eq:fp}
    \log M_*\ {\rm (AEB)} = (0.77 \pm 0.02) \log \mdust \\ 
    - (0.77 \pm 0.01) \log \rm{sSFR} - (3.1 \pm 0.2)
\end{multline}

\begin{figure*}
    \centering
    \includegraphics{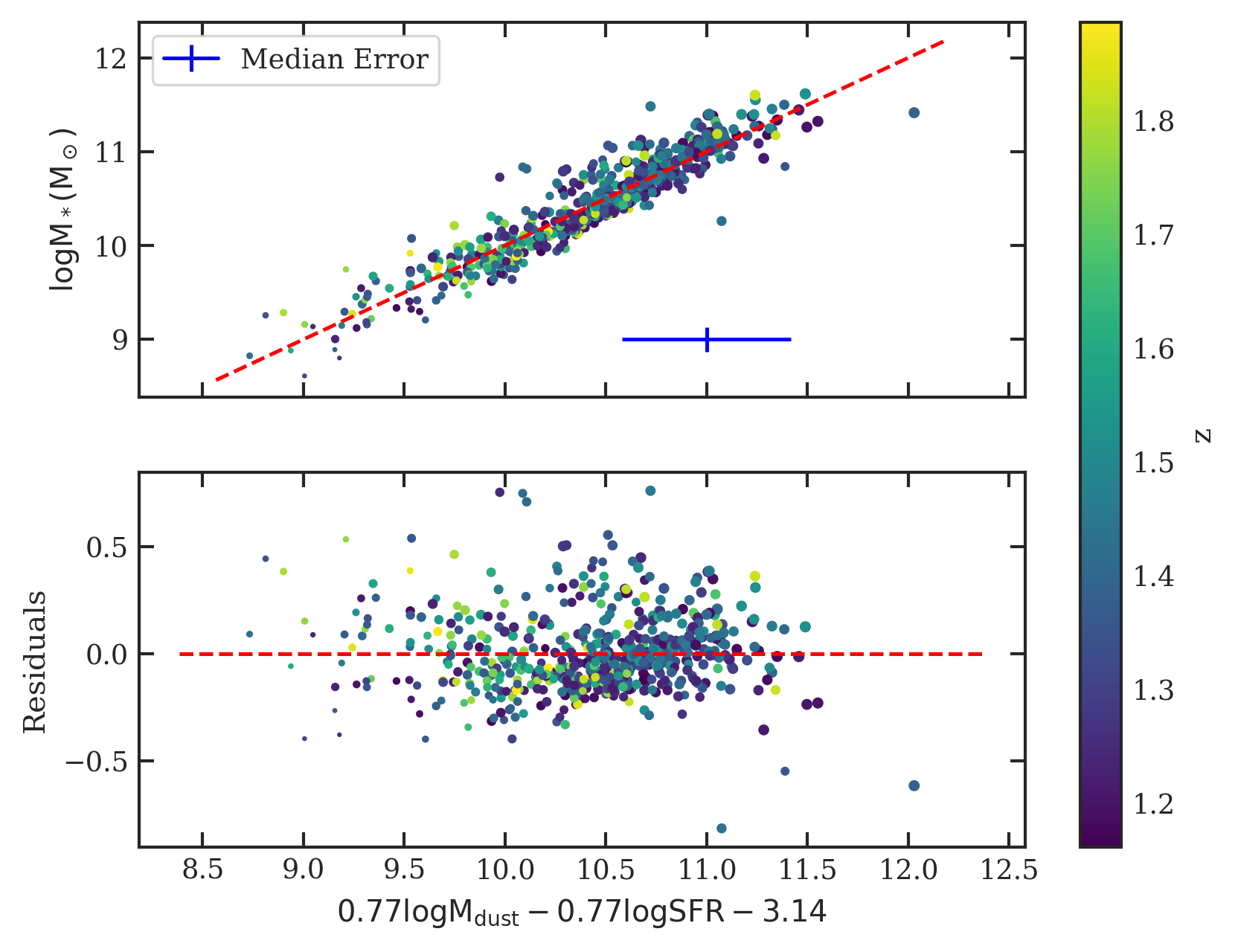}
    \caption{``SMD'' relation for sSFR, $M_*$, and \mdust (top) and residuals (bottom) when energy balance is assumed. The red line in the top panel shows the one-to-one correspondence. While some of the non-uniformity of the residuals are clearly due to the non-uniformity of stellar masses in the sample, it is still clear that there are some systematics to this relation, which is unsurprising considering only 87\% of the variance was explained with this model.}
    \label{fig:fp}
\end{figure*}

This relation is shown in Figure \ref{fig:fp}. The model has a tight fit ($R^2=0.87$), with a quoted residual standard error of $0.18$ dex, although given that the three parameters are not independent, this is not quite a measure of the true scatter in the relations. 

When we let dust mass be a free parameter, as discussed earlier, there is considerably more scatter. We found that the sources that produced the most scatter are the faintest in our sample, so we restrict our analysis to $\jhmag<24$.  (This removes 33 out of the 603 non-AGN from the dataset.) With this restriction, the best-fit relation becomes
\begin{multline} \label{eq:fpdmf}
    \log M_*\ {\rm (DMF)} = (0.25 \pm 0.02) \log \mdust \\ 
    - (0.58 \pm 0.02) \log \rm{sSFR} - (3.0 \pm 0.2)
\end{multline}

\begin{figure*}
    \centering
    \includegraphics{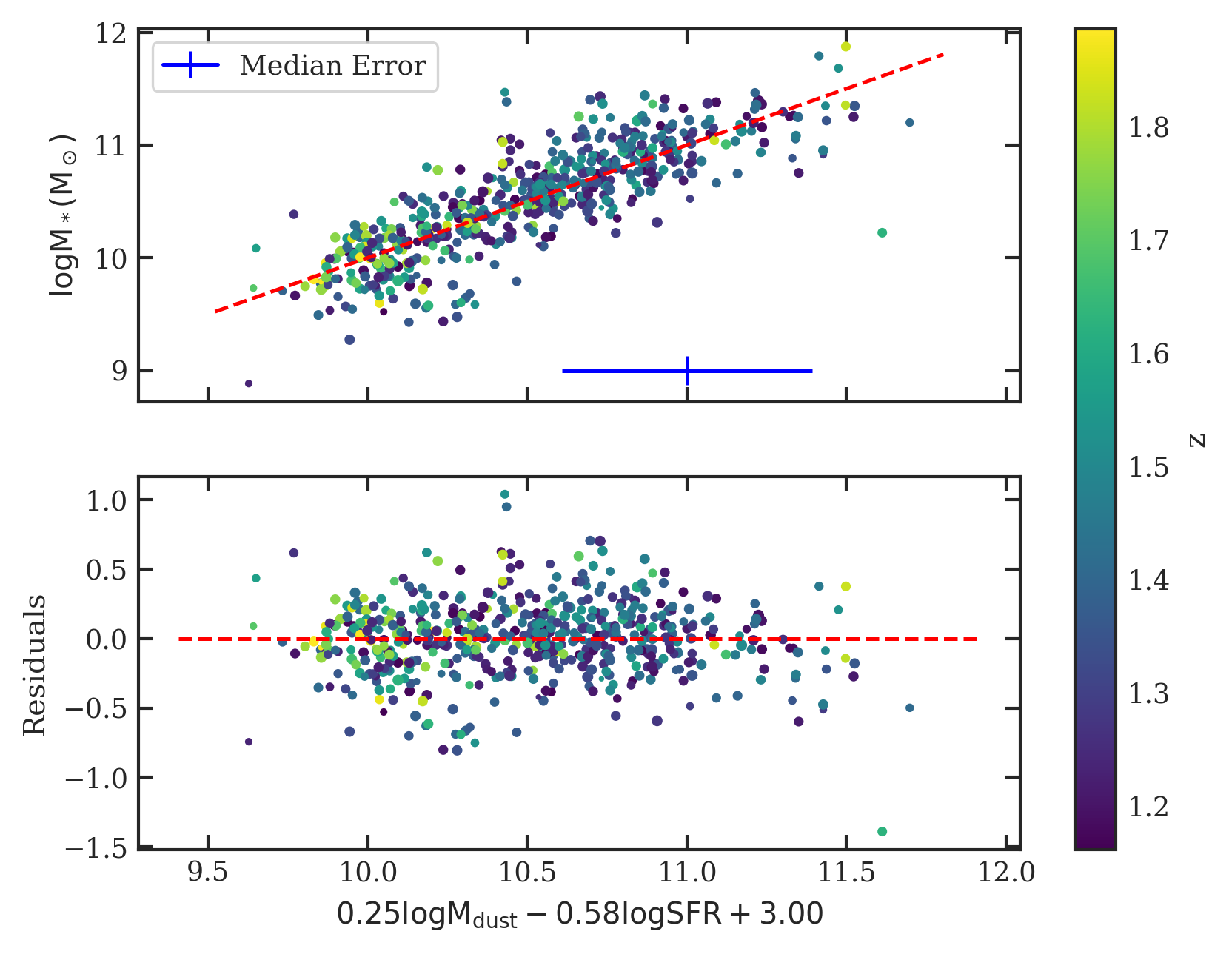}
    \caption{``SMD'' relation for sSFR, $M_*$, and \mdust (top) and residuals (bottom) when dust mass is left free. The red line in the top panel shows the one-to-one correspondence. This relation is quite different from that found for the case where energy balance is assumed. In addition, the scatter is quite reasonably much larger, and there is a much lower $R^2$ value of $0.69$, with a root-mean-square error of $0.26$ dex}
    \label{fig:fpdmf}
\end{figure*}

This relation is shown in Figure \ref{fig:fpdmf}. The model is not nearly as successful as Equation \ref{eq:fp} ($R^2=0.69$ vs.\  $R^2=0.87$), with a quoted residual standard error of $0.27$ dex. In this case, the dust mass measurements are mostly independent of stellar mass and sSFR, so the error measurement is more realistic than the one for Equation \ref{eq:fp}.

\begin{figure*}
    \centering
    \includegraphics{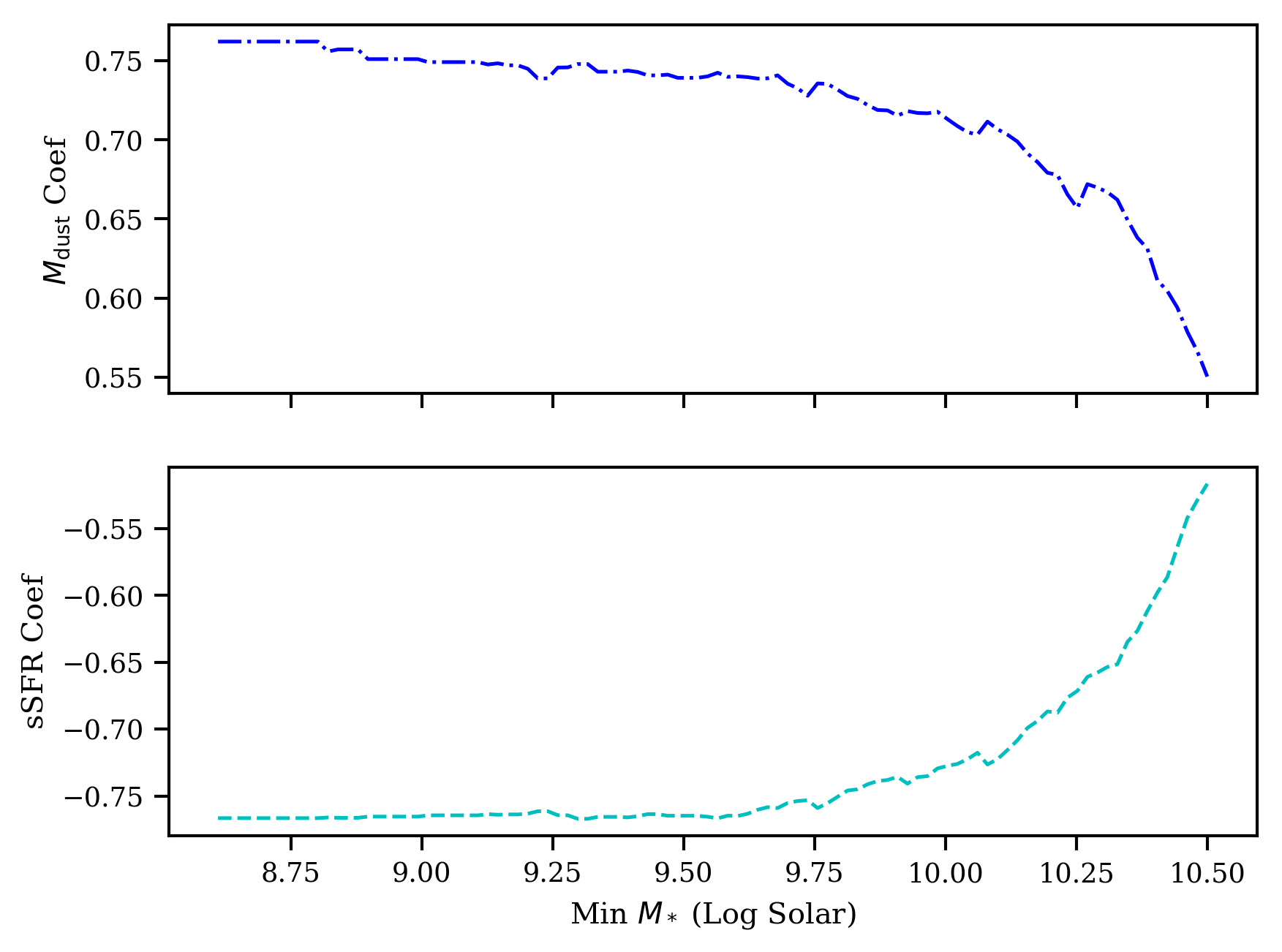}
    \caption{Coefficients of $\log \mdust$ and $\log \rm{sSFR}$ for the minimum sliding test with $\log M_*$ as the dependent variable when energy balance is assumed. While the coefficients rapidly change when the minimum of the stellar mass distribution is raised above $\log M/M_{\odot} \sim 10$, the relationship is stable for lower values. Given the predicted flux limits of near-future missions, our results should hold for the expected catalogs of sources.}
    \label{fig:aebfpmstarmin}
\end{figure*}

In order to test for the effects of incompleteness and bias on the relationship, we used the following test: we let each of the three variables in the relation ($\log M_*$, $\log \mdust$, and $\log \rm{sSFR}$) be the dependent variable in a multivariate linear relationship. For each case, we varied the minimum (maximum) value of that dependent variable from the minimum (maximum) in the data set to a value closer to the median value, in order to simulate incompleteness. For example, when stellar mass was the dependent variable, we varied the minimum $\log M_*$ from $8.61$ to $10.5$ over 100 intervals. By observing the changes in the coefficients, we can examine how stable the overall relationship is to potential incompleteness in the parameters. Figure \ref{fig:aebfpmstarmin} shows the results of the ``minimum sliding test'' for $\log M_*$ when energy balance is assumed.

In all cases (including Figure \ref{fig:aebfpmstarmin}), changing the range of the dependent variable significantly alters the fitted coefficients. One thing to note is that the end behaviors (close to full sample inclusion) are convergent, suggesting some stability in the coefficients of Equation \ref{eq:fp}. However, there are few sources, for example, with very low or high stellar masses in our sample, meaning such regions of parameter space do not affect the overall fit greatly. Perhaps convergence would be slower if we pushed to lower flux limits and identified lower mass galaxies, thus implying that the reported coefficients are a result of the current mass distribution. 

Nevertheless, in near-future missions like \textit{Euclid} and
\textit{NGRST}, the vast majority of identified sources will be found under brighter flux limits than 3D-HST (e.g., Figure 5 in Paper~I), so in a sense we would expect to find the same relationship. In any case, since the assumption of energy balance also plays a large role in shaping the relation of Equation \ref{eq:fp}, perhaps we should expect to see a trend similar to that of Equation \ref{eq:fpdmf} and Figure \ref{fig:fpdmf} instead.

Looking at Equation (\ref{eq:fp}), we notice that at fixed stellar mass, the dust mass is proportional to the sSFR\null. As seen in Figure \ref{fig:aebfpmstarmin}, even after severe restrictions in the stellar masses are applied as part of the sliding minimum test, the proportionality remains: the coefficients of $\log \mdust$ and $\log \rm{sSFR}$ are always nearly equal and opposite. This is quite an interesting, especially when one considers a two-component dust model \citep[e.g.,][]{CharlotFall2000}, where dust is preferentially distributed around areas of star formation. Naively, one might think that increasing the sSFR would increase the dust content around younger stars, but leave the diffuse component relatively unchanged.  Equation (\ref{eq:fp}) suggests otherwise. 

On the other hand, when dust mass is left as a free parameter, the relationship between it and sSFR at fixed mass is quite different. Here, \mdust increases much more slowly than sSFR; this is consistent with the idea that only birth-cloud dust increases with sSFR. Of course, given the large scatter in the DMF relation (Equation \ref{eq:fpdmf}) and the large covariances produced by assuming energy balance (Equation \ref{eq:fp}), we cannot form any real conclusions.

In either case, we see that as the stellar mass increases, the coefficient for dust mass vs.\ sSFR (at fixed stellar mass) increases. We can justify this trend with the following explanation. Given the Kennicutt-Schmidt relation \citep{Kennicutt1998b}, the SFR surface density is fundamentally linked to the gas surface density, including the gas in molecular form \citep{Bigiel2008}. As such, we can think of sSFR as a direct probe of the (molecular) gas reservoirs in a galaxy. In the local universe, the dust-to-(molecular) gas ratio is found to increase with metallicity \citep{Draine2007,Leroy2011,RemyRuyer2014}, and in Paper~I we show that our sample is consistent with the mass-metallicity relationship in \cite{Erb2006}. Thus, as stellar mass increases, metallicity also tends to increase, raising the dust-to-gas ratio and therefore the coefficient for dust mass as a function of sSFR.

Given the non-uniformity of the residuals (bottom panels of Figures \ref{fig:fp} and \ref{fig:fpdmf}), there are clearly systematics to the ``SMD'' relation, whether or not energy balance is assumed.  At least some of the non-uniformity can be explained by clustering and selection effects in the data, including the relative lack of high-redshift, low-mass sources.

Most likely, though, there are more variables involved in the relation. For example, if we include the variables \umin and \gam described in \S \ref{subsec:individual}, we get a much tighter fit. For example, when energy balance is assumed, we find the following expression for stellar mass.

\begin{multline}
    \log M_* = (0.92 \pm 0.01) \log \mdust \\ 
    - (0.878 \pm 0.008) \log \rm{sSFR} + (1.8 \pm 0.1) \gam \\
    + (0.075 \pm 0.002) \umin - (5.8 \pm 0.1)
\end{multline}

In this expression, $R^2\sim 0.97$, and the quoted residual standard error is less than 0.1 dex. However, the specificity of these additional parameters to the \cite{DraineLi2007} dust model makes it less universal than the ``SMD'' Relation given by Equation (\ref{eq:fp}). In addition, the fact that all five variables are correlated complicates the picture and suggests that the $R^2$ value is not an accurate measure of the amount of scatter in stellar mass explained by the correlation.


\subsection{H$\alpha$/H$\beta$ Stacking and Nebular Attenuation}

\begin{figure}[!ht]
    \centering
    \resizebox{\hsize}{!}{
    \includegraphics{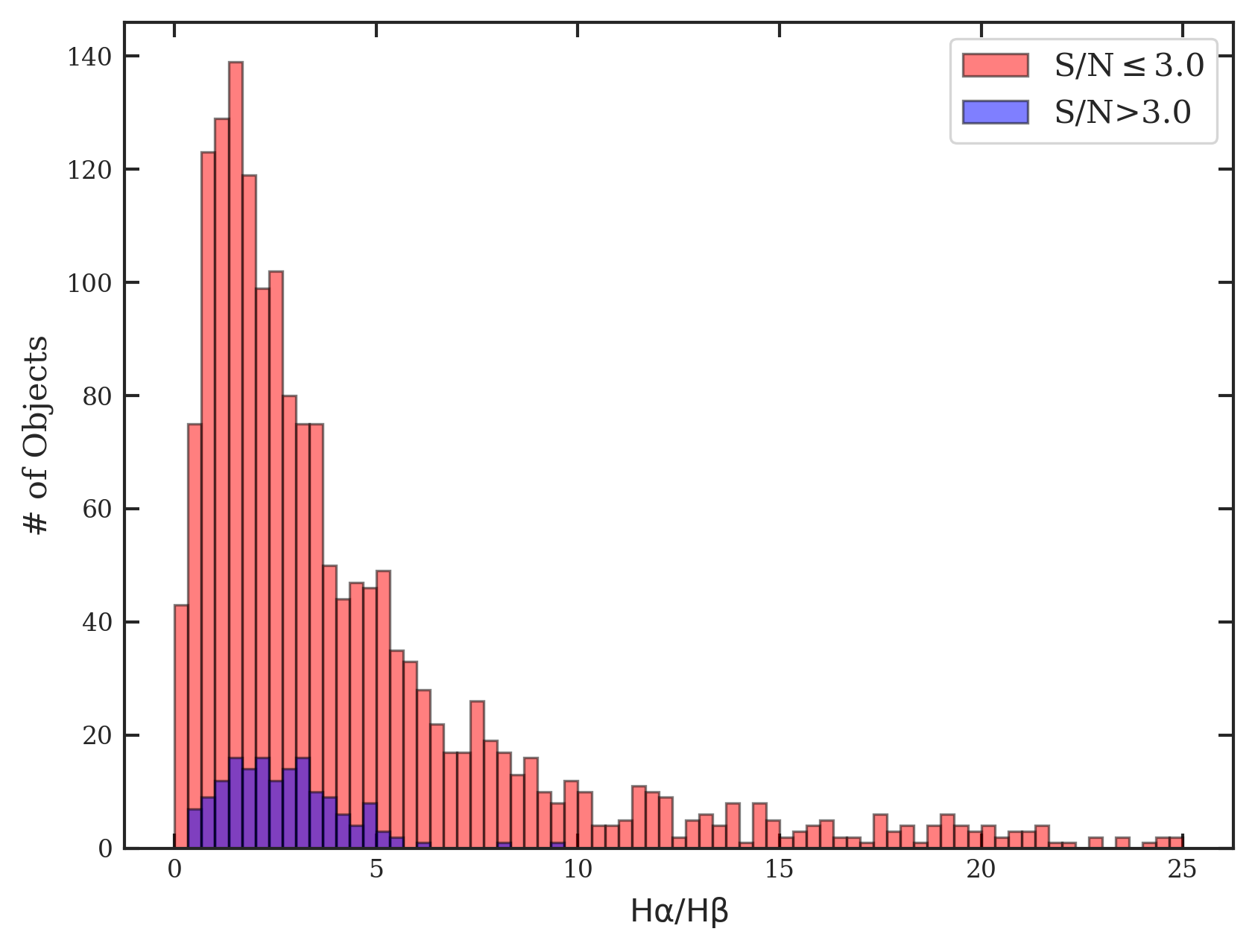}
    }
    \caption{Histogram of observed \Ha/\Hb ratios as measured by \cite{Momcheva2016} for our entire sample of $1.16 < z < 1.9$ galaxies. The low signal-to-noise ($S/N \leq 3$ for the ratio) objects are in red; galaxies with $S/N>3$ are in blue.  Even in the high signal-to-noise sources, there are a large number of sources with unphysical line ratios.}
    \label{fig:hist}
\end{figure}

For sources with redshifts $1.22<z<1.55$, \Ha and \Hb are both present on the G141 grism data. For such objects, we should be able to derive an independent measure of attenuation using the line fluxes given in the 3D-HST catalog \citep{Momcheva2016}. Unfortunately, given the low resolution of the G141 grism and the fact that the \Hb fluxes tend to be near or even below the survey's flux limit, Balmer decrement measurements for individual $1.16 < z < 1.9$ galaxies are problematic.  Figure \ref{fig:hist} illustrates this issue. Since the \Ha/\Hb ratio depends almost exclusively on atomic physics, it cannot be less than $\sim 2.86$, yet there are a significant number of sources where this is the case.  Moreover, some of the non-physical ratios occur in systems with relatively high signal-to-noise.

Hydrogen absorption from the underlying stellar population could be an issue with the Balmer line measurements, especially for the very weak \Hb line.  However, the amplitude of this absorption in the vigorously star-forming galaxies of our sample should be small, and its effect has been included in the 3D-HST line flux measurements.  

But there are other issues that could affect the observed ratio. First, the \Ha values we adopt are in fact the sum of \Ha and [\ion{N}{2}], and at the resolution of the G141 grism, these recombination and collisionally excited lines are extremely difficult to disentangle.  Of course, this issue would not create non-physical Balmer-line ratios, since the presence of [\ion{N}{2}] in the \Ha measurement could only increase the observed \Ha/\Hb ratio.  

Another factor is that at the low end of our redshift window, \Hb lies at the blue edge of the G141 grism's wavelength range, where the system sensitivity is rapidly decreasing.  In this region, the conversion between electrons and flux carries a large uncertainty. However, we do not observe a significant correlation between \Ha/\Hb and redshift, and the presence of unphysical line ratios is not limited to the low end of our redshift range.

Whatever their cause, the Balmer line fluxes recorded in the \citet{Momcheva2016} catalog prove that non-physical \Ha/\Hb ratios are inherent to the survey itself and cannot be corrected in a methodical fashion. We can, however, attempt to draw conclusions about the systematics of the galaxy population using stacked spectra. 


Before using \Ha, we corrected for contamination by the [\ion{N}{2}] doublet using a prescription similar to that adopted by \cite{Price2014}. Briefly, using the stellar mass measurements from \mcsed, we fit a \cite{Tremonti2004} mass-metallicity curve lowered by 0.56 dex \citep[as befitting $z \geq 2$ star forming galaxies; see][]{Erb2006}. This curve fits our data well (see Paper~I), although there is considerable scatter. We then used the \cite{PettiniPagel2004} relation to connect metallicity to the \NII to \Ha ratio. Finally, we multiplied the \NII correction by $4/3$ to account for the contributions of \NIIl. All of these steps are quantified by

\begin{align}\label{eqn:Hacor}
    12+\log{\mathrm{O/H}} &= -2.052+1.847\log{M_*} - 0.08026\left(\log{M_*}\right)^2 \nonumber \\
    \mathrm{N2} &\equiv \log{\frac{\mathrm{NII}~\lambda 6584~\mathrm{flux}}{\mathrm{H}\alpha~\mathrm{flux}}} \nonumber \\
    \mathrm{N2} &= \frac{12+\log{\mathrm{O/H}}-8.90}{0.57} \\
    \mathrm{H}\alpha_{\mathrm{new}} \nonumber &= \mathrm{H}\alpha_{\mathrm{orig}}*\left( 1 - \frac{4}{3}10^{\mathrm{N2}} \right)
\end{align}

Formally, these corrections range from a few percent to 40\% of a given \Ha flux. However, the high [\ion{O}{3}]/\Hb ratios in our sample suggest that the ionization parameters in $1.2 < z < 1.9$ emission-line galaxies are quite high ($\log U \sim -2.5$).  Since N$^+$ has a lower ionization potential than O$^+$, we expect the \NIIt fluxes should be low as well. Thus, it is quite possible that we are overestimating the \NIIsimp~correction for our higher-stellar-mass galaxies, and therefore underestimating the nebular attenuation.

For the stacking process, we divided the galaxy sample into bins by properties such as $z$, SFR, $M_*$, \jhmag, etc. Rather than measuring the total \Ha and \Hb fluxes in the stacked bins and dividing to get the Balmer decrements, we added the already existing individual measurements for \Ha and \Hb given by \cite{Momcheva2016}, and, in each bin, we took the ratio of the resulting sums as the Balmer decrement. The reason for this decision is that the 3D-HST team used a variety of sophisticated procedures to measure the grism's 2-D line fluxes as accurately as possible. Indeed, our efforts to measure Balmer decrements from stacked 1-D and 2-D spectra resulted in more frequent nonphysical results (Balmer decrements less than 2.86) than the process outlined above using \cite{Momcheva2016} fluxes.


In our stacking analysis, we considered only sources where both \Hb and \Ha were between 1.13 and 1.65~\um in the observed frame ($1.32<z<1.51$), as this is the range in which the instrument sensitivity is relatively stable \citep[as observed by][]{Price2014}. Furthermore, sources were included only if \Ha and \Hb were located in areas with continuous grism coverage and decent continuum characterization (S/N$>0.1$ per pixel). Of such sources, we included only those for which \Ha and \Hb fluxes as well as median 5500-6000 \AA~continuum were positive and the ratio of \Ha to \Hb was at most one standard deviation under $2.86$. The value $2.86$ is used as the intrinsic \Ha/\Hb ratio from basic atomic physics \citep[and loosely from typical conditions;][]{OsterbrockFerland2006}.

We found that 164 sources matched the criteria described in the previous paragraph. We then binned the sources in such a way that each bin received roughly the same number of sources.   This meant placing $\sim 33$ objects in each of 5 bins.


While this process ensures that the individual measurements in each stack are consistent with physics, it likely biases the results toward higher extinction values. Indeed, we repeated the same process but with no minimum \Ha/\Hb requirement and found considerably lower extinction values but similar trends. For this reason, we consider the absolute values of E(B-V) to possibly be overestimates, whereas the trends (derivatives) with properties such as stellar mass and SFR are more useful for inference.

To determine our uncertainties, we performed a bootstrap analysis on the normalized sums of \Ha and \Hb with 100 iterations and used the standard deviation of the resulting ratios. To account for errors of individual measurements, we added numbers drawn from a normal distribution with $\mu=0$ and $\sigma$ given by \cite{Momcheva2016} to each \Ha and \Hb measurement selected in each bootstrap iteration.

The nebular attenuation E(B-V) can be computed directly from the Balmer decrement given a reddening law and the intrinsic ratio of 2.86. Here, we use the \cite{CCM1989} law, which results in 
\begin{equation}
    \mathrm{E(B-V)} = 2.33\log_{10}\frac{\left(\mathrm{H}\alpha/\mathrm{H}\beta\right)_{\mathrm{obs}}}{2.86}
\end{equation}
As the choice of reddening law primarily affects the coefficient ($2.33$ for Cardelli, $1.97$ for Calzetti, etc.), the trends we find should be valid for other laws as well.

\begin{figure*}
    \centering
    \resizebox{\hsize}{!}{
    \includegraphics{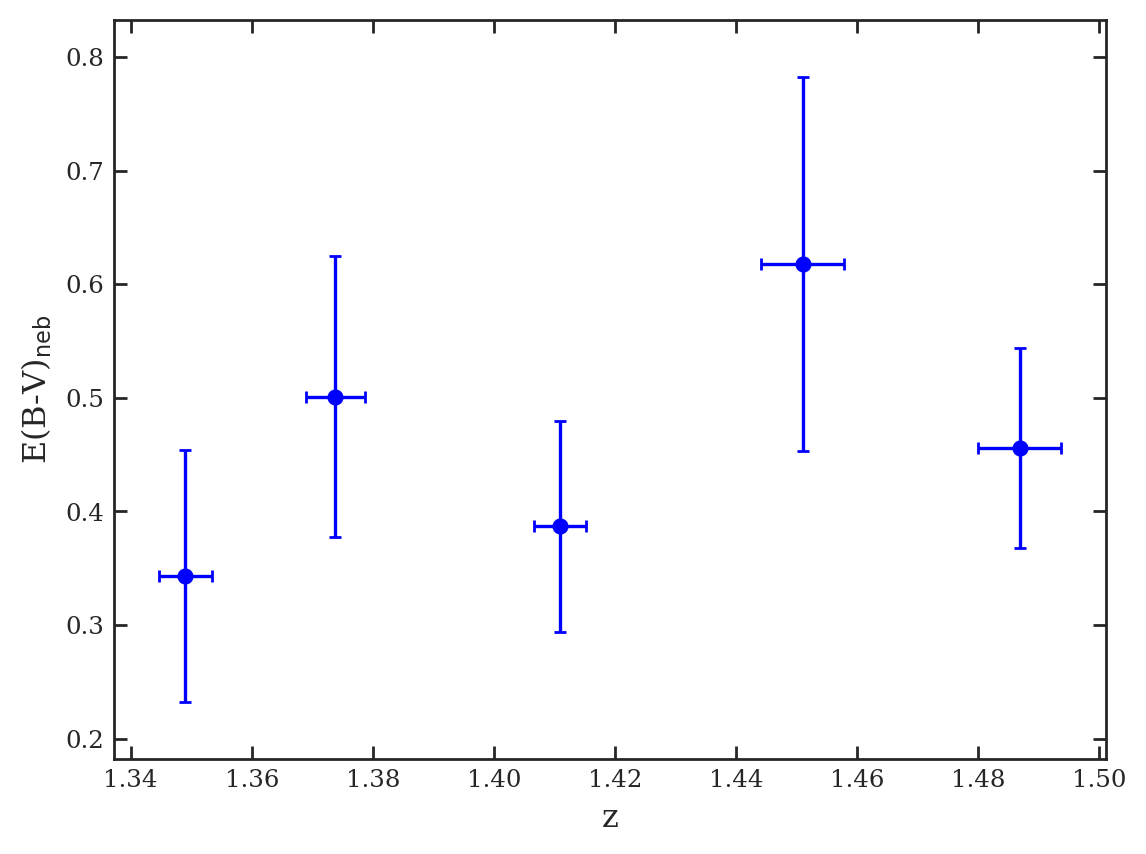}
    \includegraphics{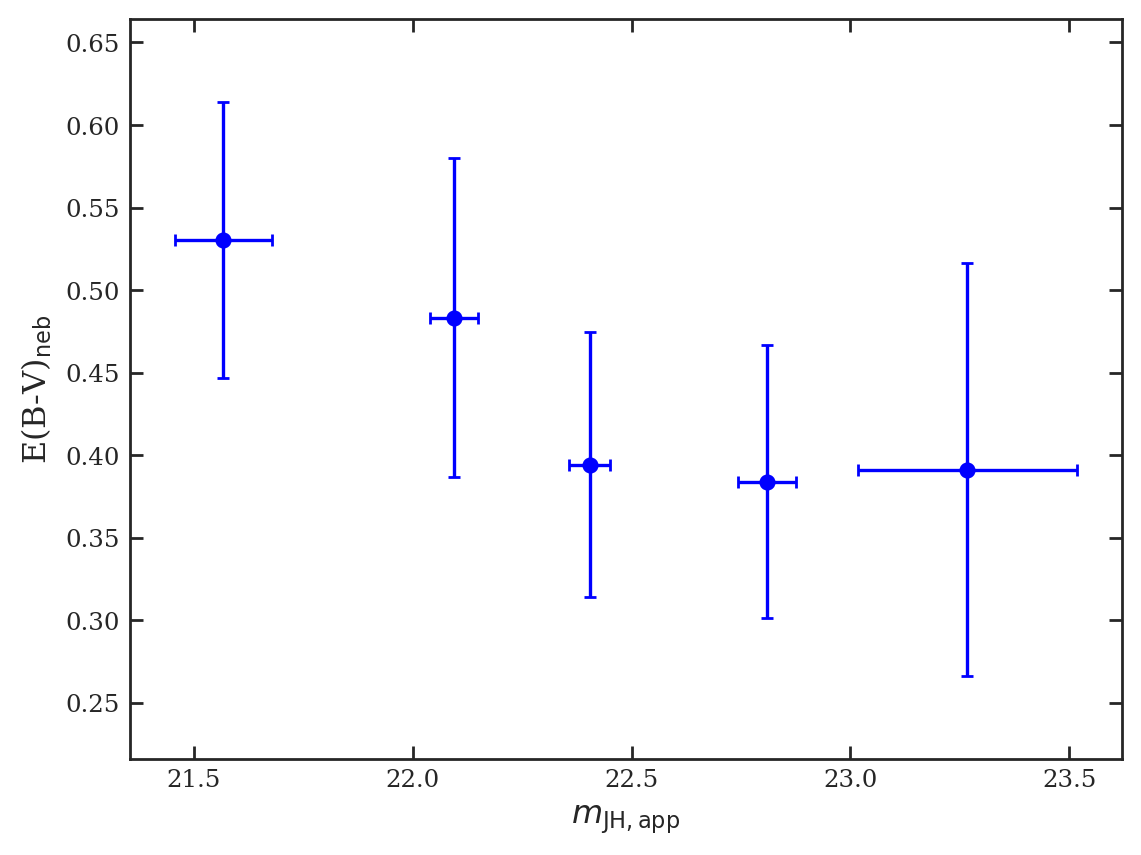}}
    \resizebox{\hsize}{!}{
    \includegraphics{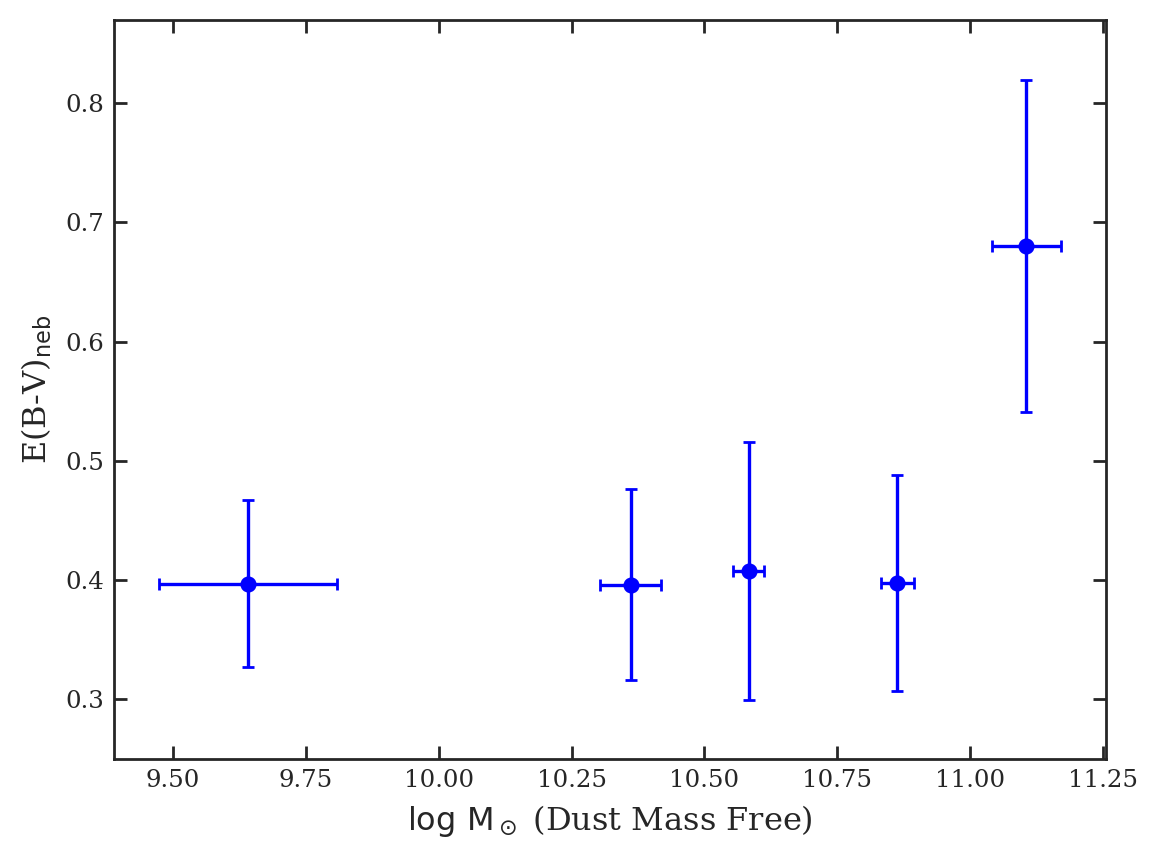}
    \includegraphics{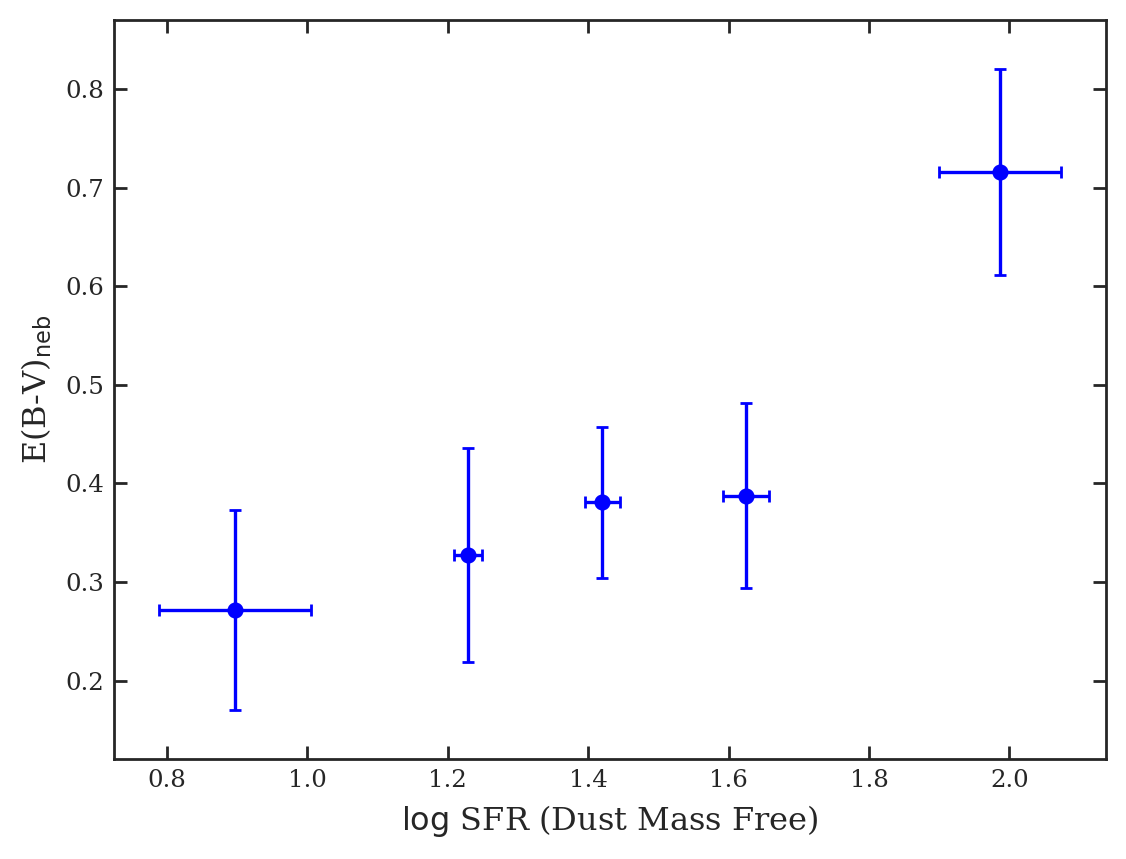}}
    \resizebox{\hsize}{!}{
    \includegraphics{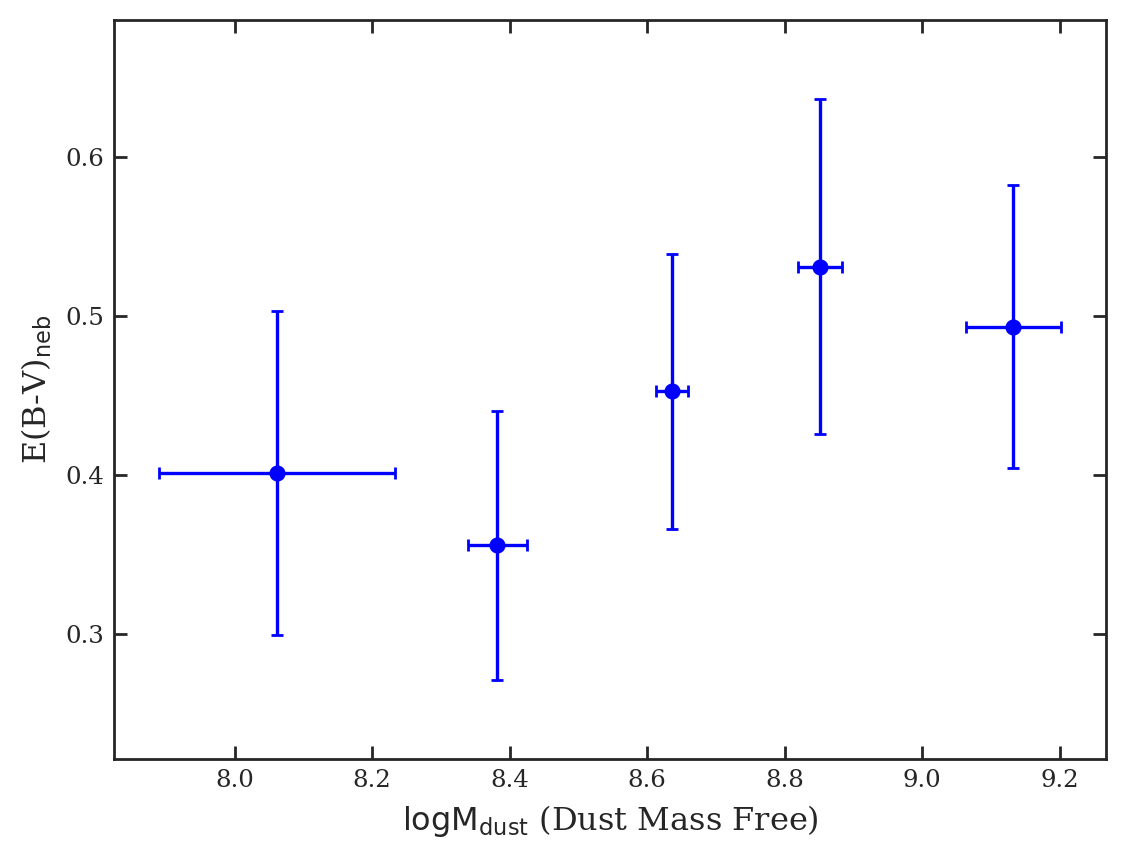}
    \includegraphics{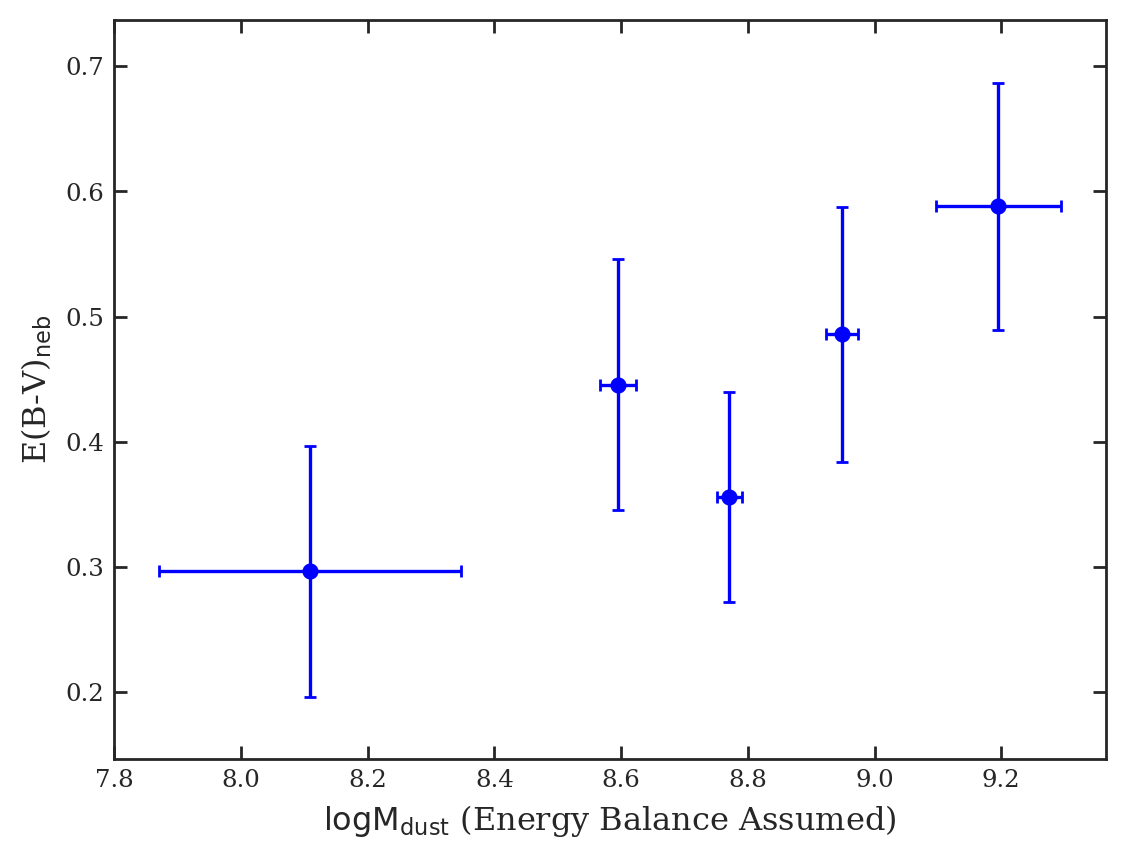}}
    \caption{Nebular attenuation E(B-V) from stacked Balmer decrements. From left to right, top to bottom, we show the behavior of E(B-V) with redshift, apparent JH magnitude, stellar mass, SFR, and dust mass. The nebular attenuation generally increases with redshift and apparent brightness. For stellar mass and SFR, we show the results for SED fitting without the energy balance assumption. We see positive trends in E(B-V) in both cases. For dust mass, we show the results for energy balance assumed (right panel) and not assumed (left panel). When dust mass is a free parameter, it is essentially independent of the nebular attenuation, but when energy balance is assumed, there is a clear positive correlation.}
    \label{fig:HaHbcor}
\end{figure*}

In Figure \ref{fig:HaHbcor}, we show the nebular attenuation E(B-V) as a function of redshift (top left), apparent JH magnitude (top right), stellar mass (middle left), SFR (middle right), and dust mass (bottom). Stellar mass, SFR, and dust mass in the bottom left panel have been computed using dust mass as a free parameter whereas dust mass in the bottom right panel has been computed with the energy balance assumption. The nebular attenuation rises monotonically or near-monotonically with stellar mass and SFR\null. No definitive comment can be made on the relationship between nebular attenuation and redshift and apparent magnitude. While attenuation seems to rise with redshift and increasing brightness, the large error bars mean that our results are also consistent with no correlation.

\begin{figure}
    \centering
    \resizebox{\hsize}{!}{
    \includegraphics{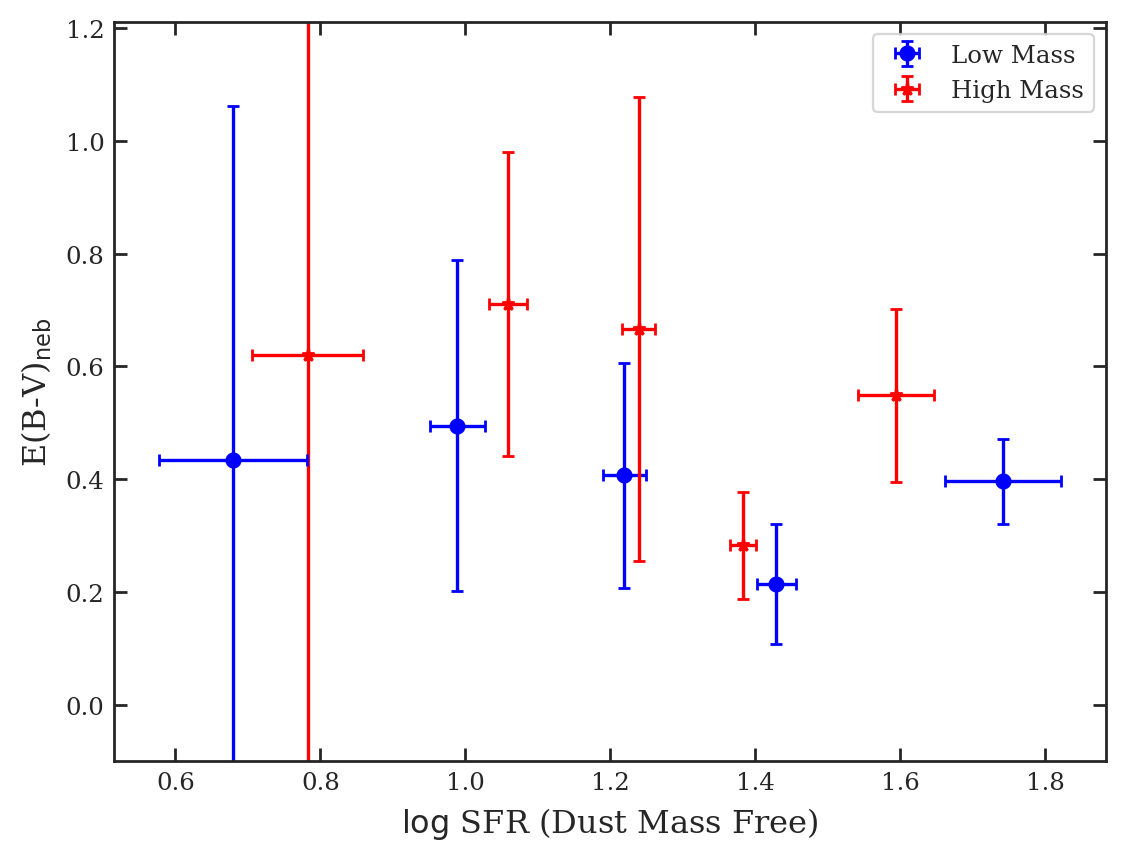}}
    \caption{Nebular attenuation E(B-V) from stacks in SFR. Here, we divide the sample into a lower-mass sample and an upper-mass sample (with the median mass as the divider). We find no evidence for a correlation between the attenuation and SFR in either mass-divided sample, suggesting that the correlation between stellar mass and SFR (i.e., star forming main sequence) is stronger than the relation between nebular extinction and SFR.}
    \label{fig:HaHbSFRMass}
\end{figure}

On the star forming main sequence \citep[e.g.,][]{Brinchmann2004,Noeske2007} stellar mass and SFR increase concurrently. At the same time, star formation and evolution tend to create more dust, so an increase in E(B-V) with SFR and stellar mass is unsurprising, and is consistent with the literature \citep[e.g.,][]{Pannella2015,Bogdanoska2020}. In Figure \ref{fig:HaHbSFRMass} we test the relative strengths of the correlation between stellar mass, SFR, and  dividing each SFR stack into a low-mass and high-mass bin.  The result of this procedure is that the correlation between attenuation and SFR disappears, suggesting that the stellar mass -- SFR relation is stronger. A caveat of this result is that dividing the already small bins (33 objects each) in half pushes us squarely into the realm of small-number statistics, so a larger sample would be needed to make a more statistically sound conclusion.

Given that the mass distribution of our galaxies is relatively uniform with redshift, the (tentative) increase in E(B-V) with redshift suggests that galaxies tend to be dustier as we peer farther into the early universe. The picture here may be convoluted by the greater impact of flux limits on the accessible high-$z$ parameter space (and the large uncertainties in the nebular attenuation).

As for dust mass, we find stronger evidence for a positive correlation between E(B-V) and dust mass when energy balance is assumed than when dust mass is free.  This makes sense as in the former case the measurement of dust mass is intricately connected to the UV-NIR spectrum. When dust mass is free, our results are consistent with both a positive correlation and no correlation between the two parameters. While we do expect a positive trend since more dust naturally leads to more attenuation, we can expect the star-dust geometry to create significant amount of scatter, as discussed in \S \ref{subsec:dmfvsaeb}.



Of course, a more detailed analysis involving the morphological characteristics of the galaxies with higher signal-to-noise Balmer decrements is required to gain a more complete understanding of the properties of nebular attenuation. A larger sample in general would help increase the statistical significance of the results and hopefully shrink the error bars.

\section{Conclusion}\label{sec:disc}

Dust is a major player in galaxy evolution, affecting star formation and chemistry and providing up to 30\% of galaxies' bolometric outputs. While dust has been extensively studied in the local universe, relations among dust, gas, and radiation at high redshift have not yet been well defined due to the small number of  photometric and spectroscopic observations.  One source of potential high-$z$ targets are IR grism-selected emission-line galaxies. While ELGs are subject to certain biases and incompleteness, emission-line surveys are an efficient way to identify large samples of high-$z$ galaxies with accurate redshifts.  Millions of such galaxies will soon be found by the upcoming \textit{Euclid} and the \textit{NGRST} missions.

As described in Paper~I, we have vetted a sample of 9,341 candidates from the 3D-HST grism survey \citep{Brammer2012,Momcheva2016} to create a clean sample of 4,350 ELGs with continuum magnitudes $\jhmag=26$ and redshifts $1.16<z<1.90$. Of these, 669 ELGs have mid- and/or far-IR data from \textit{Spitzer} and \textit{Herschel}. After using X-ray identifications and the \cite{Donley2012} IRAC (3.6-8.0 \um, observed-frame) criteria to remove AGN from our sample, we amalgamated the galaxies' emission line strengths with UV-FIR photometry. We then defined the galaxies' spectral energy distributions and fit their full-spectrum SEDs with mid- and far-IR data using \mcsed \citep{Bowman2020}, a fast, flexible MCMC-based code that combines emission from stars with attenuation and emissivity from dust to estimate the galaxies' stellar masses, SFRs, internal reddening, and dust mass. 

\mcsed generally yields reasonable SED fits for objects in our sample, although a few parameters tend to be poorly constrained. We investigated the assumption of energy balance, which equates the energy attenuated in the UV and optical to the energy re-emitted in the MIR/FIR, by performing two separate \mcsed fits, one requiring requiring the UV absorption to balance the IR emission, and one allowing the program to calculate an independent normalization to the dust emission curve.

While both options often lead to similar SED fits with comparable goodness-of-fit measurements and give consistent results for stellar mass, the comparison between \mcsed's SFRs and equivalent values derived using various locally calibrated SFR indicators \citep[including the FUV, NUV, \Ha, and 24 \um emission;][and references therein]{Kennicutt2012} are quite different. More specifically, the assumption of energy balance reduces the scatter in the SFR comparisons and makes the SED SFRs more compatible with those from the wavelength-specific measurements.  However, this result may simply reflect the fact that energy balance is built into the other indicators, whether in the form of attenuation corrections  (for the FUV, NUV, and \Ha) or the development of the indicator itself (24 \um, total IR).

When we directly measure the dust attenuated and emitted fluxes in the case where energy balance is not assumed, we find broad agreement but significant scatter ($\sim 0.54$ dex), casting doubt on the legitimacy of the energy balance argument for individual objects.

While we generally lack enough rest-frame FUV photometry to properly constrain the UV slope, $\beta$, we were still able to use \mcsed to investigate the ratio total FIR to UV luminosity, IRX, in our set of IR-bright galaxies.  Our IRX-$\beta$ plot shows a great deal of scatter with a best-fit curve that is generally above the \cite{Meurer1999} relation.  This suggests the presence of complex geometries and low UV optical depths in our galaxies, and possibly steeper attenuation laws compared to local starburst galaxies \citep{Narayanan2018}.

We found simple linear relations between sSFR, stellar mass, and dust mass. The assumption of energy balance leads to a much tighter relation, due to the added covariance between dust mass and the other two parameters. Equation \ref{eq:fp} and Figure \ref{fig:fp} describe a strong correlation ($R^2=0.87$) between the three quantities, with the standard error for the scatter being only 0.18 dex. While individual residuals show systematics, the relation is quite strong and only moderately affected by incompleteness in our data.

When we do not assume energy balance, the lower covariances between dust mass, stellar mass, and sSFR result in a less biased relationship with more realistic errors. Equation \ref{eq:fpdmf} and Figure \ref{fig:fpdmf} show the best-fit relation when dust mass is a free parameter. Not only are the coefficients for sSFR and dust mass significantly different than in Equation \ref{eq:fp}, but there is also much more scatter, with the standard error being 0.27 dex.

Our analysis suggests that,
while the energy balance argument is well motivated and ubiquitous in the literature, its application to individual galaxies is dubious. Furthermore, the non-negligible covariances between stellar mass, sSFR, and dust mass measurements from SED fitting codes are complex and difficult to account for statistically. From this work, it is clear that a relationship between dust mass, stellar mass, and sSFR exists, but the exact form it takes will need further study, preferably with a larger data set and hierarchical Bayesian techniques.


Because of the low signal-to-noise of \Hb measurements in individual galaxies, most of the Balmer decrements estimated for our ELGs are unreliable. However, we can still explore the systematics of the \Ha/\Hb ratio by stacking the data using various galaxy properties.  This procedure shows that nebular attenuation E(B-V) increases monotonically with stellar mass and SFR, although the relationship between stellar mass and SFR, i.e., the star-forming main sequence, seems to be stronger than that between attenuation and SFR.

We find no definitive evidence for a correlation between nebular attenuation and other properties, although our data suggest that attenuation may increase with redshift, apparent brightness, and dust mass. Future work including morphological studies of the galaxies, better grism or spectroscopic data, and measurements of metallicity will better flesh out the nature of nebular attenuation in high-redshift ELGs.

\acknowledgments

We thank the anonymous referee for their very thoughtful review, which led to a more sensible and reliable scope for the conclusions we reached through our analyses.

We thank Greg Zeimann, Joel Leja, and Alex Belles for useful conversations about methods and results, especially in the bigger context of galaxy evolution.

This work has made use of the Rainbow Cosmological Surveys Database, which is operated by the Centro de Astrobiolog{\'i}a (CAB/INTA), partnered with the University of California Observatories at Santa Cruz (UCO/Lick,UCSC). This work is based on observations taken by the CANDELS Multi-Cycle Treasury Program with the NASA/ESA HST, which is operated by the Association of Universities for Research in Astronomy, Inc., under NASA contract NAS5-26555. This research has made use of NASA’s Astrophysics Data System. This research has made use of the SVO Filter Profile Service (http://svo2.cab.inta-csic.es/theory/fps/) supported from the Spanish MINECO through grant AYA2017-84089. 

Computations for this research were performed on the Pennsylvania State University’s Institute for Computational and Data Sciences’ Roar supercomputer. The Institute for Gravitation and the Cosmos is supported by the Eberly College of Science and the Office of the Senior Vice President for Research at the Pennsylvania State University.

This material is based upon work supported by the National Science Foundation Graduate Research Fellowship under Grant No. DGE1255832. 

\nocite{Rodrigo2012,Rodrigo2020}

%

\vspace{5mm}
\facilities{HST (WFC3), Spitzer (MIPS), Herschel (PACS, SPIRE), GALEX, Swift(UVOT)}


\software{NumPy \citep{Numpy}, AstroPy \citep{Astropy2013,Astropy2018}, SciPy \citep{Scipy2001,Scipy2020}, CLOUDY \citep{Ferland1998,Ferland2013}, FSPS \citep{Conroy2009,Conroy2010SPSM}, \mcsed \citep{Bowman2020}, R \citep{R}}




\bibliography{sample63}{}
\bibliographystyle{aasjournal_mod}



\end{document}